\documentclass[twocolumn]{aastex701}

\usepackage{amsmath}

\begin{document}

\title{Angular Momentum Transport across Accretion States around a rapidly spinning black hole}

\author[0009-0002-2354-2884]{Rohan Raha}
\affiliation{Department of Physics, Indian Institute of Science, 
Bangalore 560012, India}
\email{raharohan@iisc.ac.in}

\author[0000-0002-2825-3590]{Koushik Chatterjee}
\affiliation{Canadian Institute for Theoretical Astrophysics, University of Toronto, 60 St. George Street, Toronto, ON M5S 3H8}
\email{kchatt@cita.utoronto.ca}

\shorttitle{Angular Momentum Flux Transport in Accretion Disks}
\shortauthors{Raha and Chatterjee}

\begin{abstract}
Accretion requires the outward transport of angular momentum, but the relative contributions of matter advection, turbulent stresses, ordered magnetic fields, winds, and black-hole spin remain uncertain in strongly magnetized flows. We use high-resolution three-dimensional GRMHD simulations around a rapidly rotating black hole ($a=0.998$) to compare SANE, intermediate-flux, and magnetically arrested disk states. In SANE and intermediate states, outward stresses are insufficient to overcome inward angular-momentum advection, and the net flux remains inward. In MAD, ordered Maxwell stresses regulate the disk and drive winds, while the Blandford–Znajek jet carries a distinct, near-constant flux extracted from the black hole, making the total angular-momentum flux outward. Magnetic-flux eruptions modulate both channels. During an eruption, an equatorial current sheet extends beyond the ergosphere, and low-density, magnetically dominated bubbles of expelled flux enter the polar region. The jet becomes asymmetric, its tilt more than doubles, and its opening angle first widens and then contracts below its pre-eruption value. Eruptions suppress the horizon accretion rate and briefly reverse the angular-momentum flux through the jet surface, while reorganizing the Maxwell stresses in the disk. As horizon-threading flux re-accumulates, the turbulent stress recovers toward its saturated MAD level. Eruption durations and lags among horizon flux, accretion rate, and jet tilt of $\sim40$--$90\,r_g/c$, correspond to milliseconds for stellar-mass black holes and tens of minutes for Sgr\,A$^*$. These results distinguish disk-regulating magnetic torques from black-hole-sourced jet transport and show how flux eruptions couple disk accretion, winds, and jet morphology.
\end{abstract}


\section{Introduction}

Accretion onto black holes powers some of the universe's most energetic phenomena, from stellar-mass X-ray binaries exhibiting dramatic state transitions \citep{Remillard2006, Fender2004} to supermassive black holes launching relativistic jets to megaparsec scales \citep{Blandford2019}. The fundamental challenge is understanding how angular momentum is transported outward to enable inward mass flow \citep{Pringle1981}. While \citet{Shakura1973} parameterized this transport through a dimensionless viscosity parameter $\alpha$, the physical origin remained unclear until \citet{BalbusHawley1991} demonstrated that the magnetorotational instability (MRI) generates turbulence in weakly magnetized differentially rotating disks, providing effective viscosity through Maxwell stresses and establishing magnetohydrodynamic processes as the primary angular momentum transport mechanism.

General relativistic magnetohydrodynamic (GRMHD) simulations \citep{Gammie2003, DeVilliers2003, McKinney2004} have revealed a fundamental dichotomy depending on magnetic flux threading the black hole. Systems with sufficient poloidal flux enter a magnetically arrested disk (MAD) state \citep{Narayan2003, Igumenshchev2003, Tchekhovskoy2011}, characterized by flux saturation, episodic eruptions, and powerful jets via the Blandford--Znajek mechanism \citep{BlandfordZnajek1977}, while lower-flux systems evolve into a standard and normal evolution (SANE) state \citep{Narayan2012} dominated by MRI with weak, disorganized fields. This dichotomy has been invoked to interpret spectral state transitions in X-ray binaries \citep{Raha2026,raha2026b,Begelman2015}, jet power variations in AGN \citep{Zamaninasab2014}, and Event Horizon Telescope observations of M87 and Sgr~A$^*$ \citep{EHT2019L5, EHT2022SgrAL5}.

Despite this progress, detailed angular momentum transport mechanisms remain poorly understood, particularly for rapidly spinning black holes where frame-dragging and the Blandford--Znajek mechanism fundamentally alter the physics. While previous studies characterized global MAD and SANE properties \citep{Tchekhovskoy2011, Narayan2012, Porth2019, Salas2024, raha2026b}, systematic angular momentum transport analyses focused on non-spinning or moderately spinning systems \citep{Penna2010, Hawley2011, Chatterjee2022, Manikantan2024}. \citet{Chatterjee2022} showed disk winds powered from flux eruptions remove angular momentum in non-spinning MADs, while \citet{Manikantan2024} found equal contributions from winds and turbulence in thick MADs around rapidly spinning black holes. However, these contributions represent averages over eruption cycles: the flux eruptions themselves seed the turbulence and power the winds \citep{Chatterjee2022}, while the jet primarily carries away black-hole angular momentum, coupling to the wind only near the horizon and leaving the disk's angular-momentum budget largely unaffected. Furthermore, a direct comparison among SANE, intermediate (INT), and MAD states around a rapidly spinning black hole remains lacking. Recent simulations also revealed intermediate states with partial flux saturation \citep{Raha2026, raha2026b}, yet their transport mechanisms remain unexplored. The central question is how the spin-driven horizon torque changes the source, carrier, and destination of angular momentum as the magnetic flux increases.

We investigate angular momentum transport in SANE, INT, and MAD configurations around a rapidly spinning black hole using high-resolution 3D GRMHD simulations with H-AMR \citep{Liska2022}. We decompose the total transport into advective and stress-driven components, separate the Maxwell stress into laminar and turbulent contributions, and compute fluxes through surfaces separating the jet, wind, and disk to characterize angular-momentum exchange among these zones.

Section~\ref{sec:numerical_setup} describes the numerical setup and state classification, Section~\ref{sec:disk_properties} defines the flow regions, and Section~\ref{sec:diagnostics} introduces the transport diagnostics. Section~\ref{sec:results} presents the quasi-steady angular-momentum budgets, Section~\ref{sec:flux_eruption} examines their evolution during flux eruptions, and Section~\ref{sec:conclusion} summarizes the results and implications.

\section{\label{sec:numerical_setup}Numerical Methods}

We employ the H-AMR code \citep{Liska2022} to solve the general relativistic magnetohydrodynamic (GRMHD) equations for magnetized accretion flows around rapidly spinning black holes. The simulations are performed in the Kerr spacetime with dimensionless spin parameter $a = 0.998$, using Kerr-Schild coordinates. We adopt geometric units with $c = GM = 1$ throughout.

Our simulations begin with a Fishbone-Moncrief equilibrium torus \citep{Fishbone1976} threaded by different magnetic field configurations. We employ an ideal gas equation of state with $p_{\text{gas}} = (\tilde{\Gamma} - 1)u_{\text{gas}}$, where $\tilde{\Gamma} = 13/9$ is the adiabatic index. The computational domain extends from the inner boundary at $r_{\text{in}} = 0.87r_H$ to $r_{\text{out}} = 10^6 r_g$, where $r_H$ is the horizon radius and $r_g = GM/c^2$ is the gravitational radius.

For our high-resolution 3D simulations, we utilize a grid resolution of $N_r \times N_\theta \times N_\phi = 448 \times 240 \times 192$ with logarithmic radial spacing. We de-refine the $\phi$ resolution by a factor of 2 within $30^{\circ}$ from the poles to increase code speed. This de-refinement does not compromise the resolution of the jet. In the MAD run, for a typical snapshot, the azimuthally averaged boundary of the magnetically dominated funnel ($\sigma = b^2/\rho = 1$; Section~\ref{sec:disk_properties}) lies at $\theta \approx 69^\circ$, $41^\circ$ and $22^\circ$ from the pole at $r = 5$, $20$ and $50\,r_g$, so the jet half-width is resolved by $\approx 92$, $54$ and $29$ cells in $\theta$. Near the jet base most of the funnel lies at $\theta > 30^\circ$, where the full 192 azimuthal cells are used. At $50\,r_g$, where the funnel lies entirely within the de-refined zone, the 96 azimuthal cells ($\Delta\phi = 3.75^\circ$) correspond to an arc of $\sin\theta\,\Delta\phi \approx 1.4^\circ$ at the jet boundary, about twice the $\theta$ spacing. We also enable local adaptive time-stepping, increasing speed by approximately a factor of 3 and reduces inversion errors near the pole \citep{Chatterjee:2019, Liska2022}. We apply outflowing boundary conditions at the radial boundaries, transmissive conditions at the poles, and periodic boundary conditions in the azimuthal direction.

\begin{figure}[ht]
\centering

\includegraphics[width=0.45\textwidth]{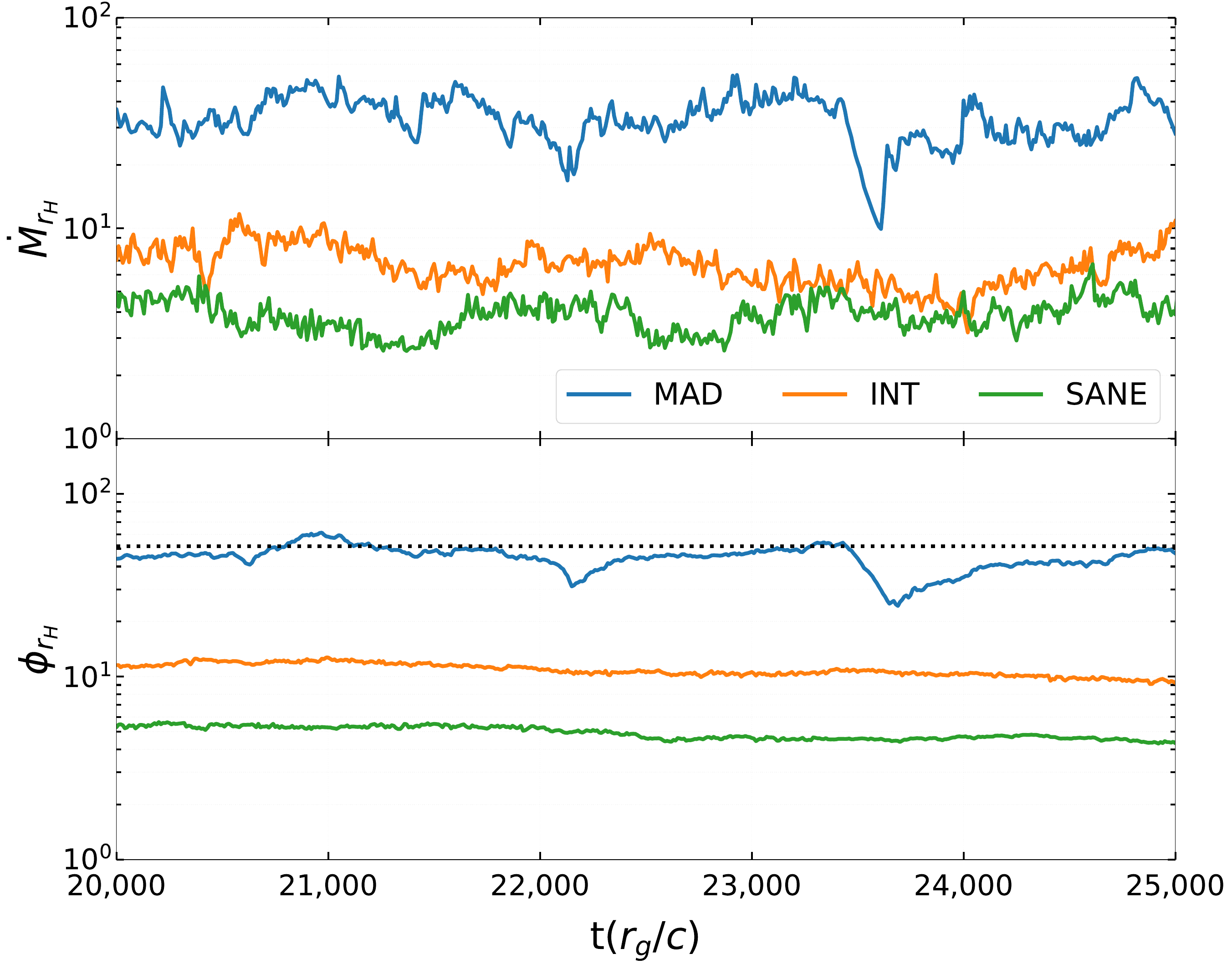}

\caption{Top: Evolution of dimensionless mass accretion rate $\dot{M}$ versus time for the three configurations. Bottom: Evolution of normalized magnetic flux $\phi$ versus time. The horizontal dashed line indicates the MAD saturation threshold $\phi_{\text{MAD}} \approx 50$.}
\label{fig:diagnostics}
\end{figure}

We focus on three representative configurations whose temporal evolution of the mass accretion rate and horizon-threading magnetic flux distinguishes the accretion states (Figure~\ref{fig:diagnostics}). The dimensionless mass accretion rate at the horizon is calculated as:

\begin{equation}
\dot{M}(r) = -\int \rho u^r \sqrt{-g} \, d\theta d\phi,
\end{equation}
where $\rho$ is the rest-mass density and $u^r$ is the radial component of 4-velocity.

The normalized magnetic flux threading the black hole, following \citet{Tchekhovskoy2011}, is defined as:

\begin{equation}
\phi = \frac{\sqrt{4\pi}}{2\sqrt{\dot{M}}} \int |B^r| \sqrt{-g} \, d\theta d\phi,
\end{equation}

where $B^r$ is the radial component of the magnetic field.

The SANE configuration evolves toward a state with minimal magnetic-flux accumulation and maintains a consistently low flux ($\phi < 10$) throughout the evolution. Its mass accretion rate shows comparatively weak variability, and its outflows are dominated by winds rather than magnetically collimated jets.

The INT configuration develops an intermediate magnetic flux and represents a transitional state between SANE and MAD. The flux reaches moderate levels ($10 < \phi < 30$) with intermediate variability, and the system produces both a jet and disk winds.

The MAD configuration accumulates the largest horizon-threading magnetic flux ($\phi>50$) and produces a powerful relativistic jet. The system exhibits strongly variable accretion and episodic magnetic-flux eruptions associated with magnetic barriers that temporarily suppress the inflow.

We initialize all configurations with their corresponding vector potentials given in \citet{raha2025} and plasma-$\beta = 100$, where $\beta$ is the ratio of gas pressure to magnetic pressure. These distinct evolutionary pathways provide a framework for comparing how magnetic-field geometry and flux accumulation modify angular-momentum transport.

\section{General Disk Properties}\label{sec:disk_properties}

We present the time-averaged structure of the three accretion configurations over the period $t = 20,000-25,000 \, r_g/c$, when all simulations have reached a quasi-steady state up to radius $40-50 \, r_g$, as we see that the mass accretion rate remains constant up to that radius \citep{Raha2026}.

\begin{figure*}[ht!]
\centering

\includegraphics[width=\textwidth]{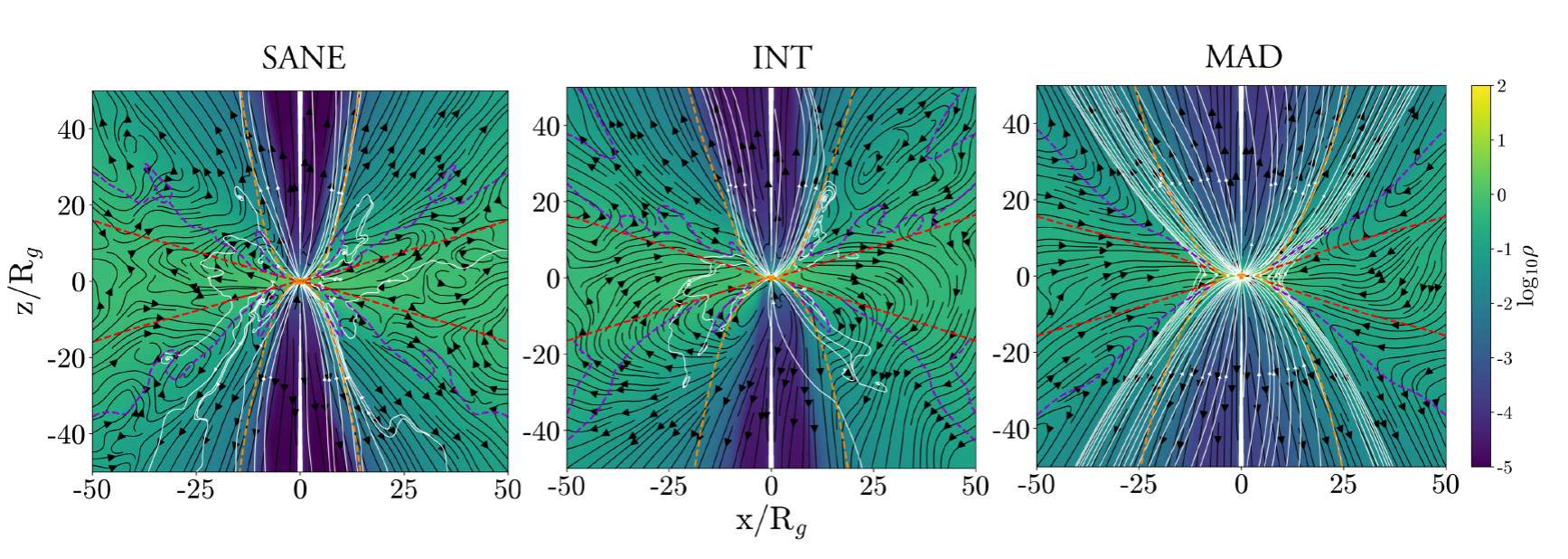}

\caption{Time-averaged density structure (color contours) with magnetic field lines (white curves) and velocity vectors (black arrows) for the three accretion configurations. Orange, red, and violet lines mark jet, disk, and wind boundaries. SANE: Tangled fields with weak coherence, an extended disk, a narrow magnetized funnel, and diffuse outflows. INT: Moderately coherent fields with mixed jet--wind morphology and transitional outflow properties. MAD: Strong poloidal fields thread the black hole, producing low-density polar funnels, a broad jet, a compact thick disk, and magnetically confined winds. The wind occupies a larger solid angle from MAD to SANE, opposite to the trend in jet extent; INT and SANE also show fragmented structures associated with infalling material.}
\label{fig:density_structure}
\end{figure*}

Figure~\ref{fig:density_structure} presents the density distribution, magnetic field topology, and velocity structure for each configuration, revealing fundamental differences in their accretion flow properties.

To quantitatively analyze the different flow components, we define three distinct regions based on physical criteria derived from the local plasma properties as shown in Figure~\ref{fig:density_structure}:

Disk Region: The disk region (red lines in Figure \ref{fig:density_structure}) is defined based on the local scale height criterion, following the approach of \citet{Porth2019}, where we identify regions within one scale height $H/r$ of the midplane. The scale height is calculated as $H/r = \langle|\theta - \pi/2|\rangle_\rho$, where  $\langle\rangle_\rho$ denotes a density-weighted average.

Jet Region: We define the jet region (delineated by orange lines in Figure~\ref{fig:density_structure}) as areas where the magnetization parameter $\sigma = b^2/\rho > 1$ \citep{raha2026b}, indicating magnetically dominated flow. This criterion captures regions where magnetic energy density exceeds the rest-mass energy density, characteristic of relativistic, magnetically driven outflows.

The jet boundary used for all angular-momentum fluxes in this work, including the time-resolved flux through the jet surface in Section~\ref{sec:flux_eruption}, is the $\sigma = 1$ contour of the time- and azimuthally averaged magnetization $\langle\sigma\rangle_{t,\varphi}$, and is therefore fixed in time. The instantaneous $\sigma > 1$ region is used only for the jet tilt and opening angle (Section~\ref{sec:flux_eruption} and Appendix~\ref{app:jet_morphology_def}); during flux eruptions it can also contain expelled magnetized material, which we identify in Section~\ref{sec:eruption_id}.

Wind Region: The wind region (violet lines in Figure \ref{fig:density_structure}) encompasses areas that lie outside the disk region and are outward-moving ($v^r > 0$), sub-relativistic, and gas-rich ($\sigma < 1$). 

However, all three configurations contain an inward-moving region adjacent to the disk that is not immediately accreted and may subsequently supply the wind. We refer to this region as the disk edge, or outer disk.

These four regions are defined hierarchically, so that they are mutually exclusive and collectively exhaustive, and the regional fluxes sum to the total flux at each radius.

\section{\label{sec:diagnostics}Diagnostics}

To analyze the mechanisms of angular momentum transport in our three accretion configurations, we examine the angular momentum flux $\dot{\mathcal{J}}$, defined following the approach of \citet{Chatterjee2022}. The total radial angular momentum flux is given by:
\begin{equation}
\dot{\mathcal{J}^i}_{\text{tot}} = \langle T^i_{\varphi} \rangle_{t,\varphi}
\end{equation}
where $\langle \rangle_{t,\varphi}$ denotes time and azimuthal averaging and $T^i_{\varphi}$ is the $i\varphi$ component of the stress-energy tensor given by 
\begin{equation}
T^{\mu\nu} = \left( \rho h + b^2 \right) u^\mu u^\nu + \left( p_{gas} + \frac{b^2}{2} \right) g^{\mu\nu} - b^\mu b^\nu ,
\label{eq:stress_energy_tensor}
\end{equation}
where $\rho$ is the rest-mass density, $p_{gas}$ is the gas pressure, $u_{gas}$ is internal energy density, $u^\mu$ is the fluid four-velocity, $h = 1 + u_{gas} + \frac{p_{gas}}{\rho}$ is the specific enthalpy, $b^\mu$ is the magnetic field four-vector in the fluid frame, $b^2 = b^\mu b_\mu$, and $g^{\mu\nu}$ is the spacetime metric. This quantity represents the net transport of angular momentum in the radial direction and can be decomposed into advective and stress-driven components as detailed in Section~\ref{sec:components}.

To understand the mechanisms driving angular momentum transport in our three accretion configurations, we analyze the radial flux of angular momentum, $\dot{\mathcal{J}^r}$, computed separately for the jet, wind, and disk regions defined by our physical criteria in Section \ref{sec:disk_properties}. The total specific angular momentum flux (normalized by the mass accretion rate at $5r_g$) is calculated by integrating the $r\varphi$ component of the stress-energy tensor (Equation~\ref{eq:stress_energy_tensor}) over the respective regions: 
\begin{equation}
\dot{\mathcal{J}}^r_{\rm region}/\dot{M}_{5r_g} = (1/\dot{M}_{5r_g}) \int_{region} \langle T^r_{\varphi}\rangle_{t,\varphi} \sqrt{-g} \, d\theta,
\label{eq:jdot_region}
\end{equation}
 where the integration domain is determined by the jet, wind, or disk conditions. 

 To understand the physical mechanisms underlying angular momentum transport in our three configurations, we decompose the total radial angular momentum flux into its constituent components following the Reynolds decomposition framework commonly used in turbulent magnetized accretion flows \citep{Balbus1998, Hawley1995, Chatterjee2022}. Any quantity $f$ can be written as $f = \langle f \rangle + \delta f$, where $\langle f \rangle$ denotes the time-averaged mean and $\delta f$ represents the fluctuation from this mean. 

The total specific angular momentum flux is decomposed into advective and stress components:
\begin{equation}
\dot{\mathcal{J}}^r = \dot{\mathcal{J}}^r_{\rm adv} + \dot{\mathcal{J}}^r_{\rm stress},
\label{eq:jdot_decomp}
\end{equation}
where the advective component represents the bulk kinetic transport of angular momentum by the mean flow:
\begin{equation}
\dot{\mathcal{J}}^r_{\rm adv} = \langle (\rho + u_{gas} + p_{gas} + b^2)  u^r \rangle \langle u_\varphi \rangle.
\label{eq:jdot_adv}
\end{equation}
This term captures angular momentum carried radially by the time-averaged radial motion of material, comprising contributions from both the gas enthalpy ($\rho + u_{gas} + p_{gas}$) and magnetic field energy density ($b^2$). 

The stress component captures angular momentum transport by correlations between velocity and magnetic field fluctuations, as well as by the mean magnetic field configuration:
\begin{equation}
\dot{\mathcal{J}}^r_{\rm stress} = \dot{\mathcal{J}}^r_{\rm Max} + \dot{\mathcal{J}}^r_{\rm Rey},
\label{eq:jdot_stress}
\end{equation}
where the Maxwell stress arises from magnetic tension and pressure forces:
\begin{equation}
\dot{\mathcal{J}}^r_{\rm Max} = -\langle b^r b_\varphi \rangle,
\label{eq:maxwell_stress}
\end{equation}
and the Reynolds stress represents turbulent velocity correlations:
\begin{align}
\dot{\mathcal{J}}^r_{\rm Rey} &= \langle (\rho + u_{gas} + p_{gas} + b^2) u^r u_\varphi \rangle \nonumber \\
&\quad - \langle(\rho + u_{gas} + p_{gas} + b^2)  u^r\rangle \langle  u_\varphi \rangle \nonumber \\
&= (\rho + u_{gas} + p_{gas} + b^2) \delta u^r \delta  u_\varphi .
\label{eq:reynolds_stress}
\end{align}
 
We further decompose the Maxwell stress into laminar (mean field) and turbulent (fluctuating field) components following \citet{Salvesen2015, Chatterjee2022}:
\begin{equation}
\dot{\mathcal{J}}^r_{\rm Max} = \dot{\mathcal{J}}^r_{\rm Max,lam} + \dot{\mathcal{J}}^r_{\rm Max,turb},
\label{eq:maxwell_decomp}
\end{equation}
where the laminar component is determined by the time-averaged magnetic field structure:
\begin{equation}
\dot{\mathcal{J}}^r_{\rm Max,lam} = -\langle b^r \rangle \langle b_\varphi \rangle,
\label{eq:maxwell_laminar}
\end{equation}
and the turbulent component captures correlations of magnetic field fluctuations:
\begin{equation}
\dot{\mathcal{J}}^r_{\rm Max,turb} = - \delta b^r \delta b_\varphi  = -\langle b^r b_\varphi \rangle + \langle b^r \rangle \langle b_\varphi \rangle.
\label{eq:maxwell_turbulent}
\end{equation}
The laminar Maxwell stress reflects organized, large-scale magnetic field configurations such as those threading the black hole in MAD states, enabling direct extraction of angular momentum via the Blandford--Znajek mechanism \citep{BlandfordZnajek1977, Tchekhovskoy2011}. The turbulent Maxwell stress, by contrast, arises from small-scale magnetic field fluctuations generated by the MRI or seeded by flux eruptions and captures the efficiency of MRI-driven turbulent transport \citep{Hawley1995, Noble2010}. The Reynolds stress captures turbulent velocity correlations and is typically subdominant in strongly magnetized regions.

In Section \ref{sec:surface_flux}, we examine the angular momentum exchange across the boundaries 
separating the jet, wind, and disk regions. This analysis reveals the pathways 
through which angular momentum is redistributed among these dynamically distinct 
zones and highlights the contrasting coupling mechanisms in the three accretion 
configurations.

For surfaces aligned with the $\hat{\theta}$ direction (such as the disk surface 
defined by constant scale height), the perpendicular angular momentum flux is 
simply $\dot{\mathcal{J}}^\theta$. However, the jet and wind boundaries form a 
tilted surface with respect to the coordinate basis. Following the approach of 
\citet{Manikantan2024}, we compute the flux perpendicular to such surfaces by 
projecting the angular momentum flux vector onto the surface normal.

For the jet surface defined implicitly by $F(r, \theta) = \sigma(r, \theta) - 1 = 0$, 
the outward normal vector in the $(r, \theta)$ basis is:
\begin{equation}
\hat{n} = \frac{1}{|\nabla F|}\left(\frac{\partial \sigma}{\partial r} \hat{r} 
+ \frac{1}{r}\frac{\partial \sigma}{\partial \theta} \hat{\theta}\right).
\end{equation}
The tangent to the $\sigma = 1$ contour makes an angle $\alpha$ with the radial 
direction, where:
\begin{equation}
\tan \alpha = \frac{r \, \partial \sigma / \partial r}{\partial \sigma / \partial \theta}.
\end{equation}
The perpendicular flux component through this tilted surface is then:
\begin{equation}
\dot{\mathcal{J}}^\perp =  \pm \dot{\mathcal{J}}^\theta \cos \alpha 
+ \dot{\mathcal{J}}^r \sin \alpha,
\label{eq:jdot_perp}
\end{equation}
where the sign depends on the hemisphere: positive for the upper hemisphere 
($\theta < \pi/2$) and negative for the lower hemisphere ($\theta > \pi/2$), 
ensuring consistent outward normal orientation. The cumulative angular momentum 
flux thus becomes 
\begin{align}
\dot{\mathcal{J}}^\perp &= 2 \pi \int^r_{r_H}  (\pm \langle T^\theta_\varphi
\rangle_{t,\varphi} |_{surface} \cos \alpha \nonumber\\
&\quad + \langle T^r_\varphi\rangle_{t,\varphi}|_{surface} \sin \alpha) \, dr',
\label{eq:jdot_cumulative}
\end{align}
where the value at any radius $r$ represents the total angular momentum that has 
crossed the surface up to that point. For the wind--disk edge interface, similar 
considerations apply when the boundary is defined by the $u^r = 0$ and $\sigma = 1$ 
condition (Section~\ref{sec:disk_properties}). Detailed component-by-component 
descriptions of the surface fluxes are provided in Appendix~\ref{app:surface_flux}, 
with Figure~\ref{fig:jdot_surfaces} therein.

To quantify the angular momentum extraction through the jet region, we compute 
the non-cumulative angular momentum flux perpendicular to the jet surface, 
following the methodology of \citet{Manikantan2024}. For a local measure of the \textit{non-cumulative} angular momentum flux through the jet surface at 
radius $r$, we calculate the local rate of angular momentum transport 
perpendicular to the jet boundary, obtained as the radial derivative of 
the cumulative flux:
\begin{align}
\dot{j}^\perp_{\rm jet\,surf}(r) = \frac{d\dot{\mathcal{J}}^\perp}{dr} &\approx  2 \pi   (\pm \langle T^\theta_\varphi\rangle_{t,\varphi} |_{\rm jet\,surf} \cos \alpha \nonumber\\
&\quad + \langle T^r_\varphi\rangle_{t,\varphi} |_{\rm jet\,surf} \sin \alpha),
\label{eq:noncumulative_jdot}
\end{align}
This quantity measures the angular momentum flux crossing the jet surface 
per unit radial extent.

In Section~\ref{sec:alpha_viscosity} we analyse the dimensionless stress-to-pressure ratio $\alpha_{\rm visc}$ \citep{Shakura1973}. Because the radial and azimuthal velocity and field components entering $\alpha_{\rm visc}$ are not individually coordinate invariant, we orthonormalize them by
$v_{\hat\varphi}=u_\varphi/\sqrt{g_{\varphi\varphi}}$, and $b_{\hat\varphi}=b_\varphi/\sqrt{g_{\varphi\varphi}}$.
 For the radial components we use $v_{\hat r}=\sqrt{g_{rr}}\,u^r$ and $b_{\hat r}=\sqrt{g_{rr}}\,b^r$, which are regular at the horizon in Kerr--Schild coordinates. With these choices every stress below is the corresponding part of $T^r_{\ \varphi}$ (Equation~\ref{eq:stress_energy_tensor}) multiplied by the common positive factor $\sqrt{g_{rr}/g_{\varphi\varphi}}$, so signs, correlations and the ratios between components do not depend on the normalization.
 
We denote by $\langle X\rangle_\rho=\int\!\!\int \rho X\sqrt{-g}\,d\theta\,d\varphi \big/ \int\!\!\int \rho\sqrt{-g}\,d\theta\,d\varphi$ the density-weighted shell average, which samples the disk body. The effective viscosity parameter is
\begin{equation}
\alpha_{\rm visc} = \alpha_{\rm Max}+\alpha_{\rm Rey},
\label{eq:alpha_decomp}
\end{equation}
with
\begin{align}
\alpha_{\rm Max} &= \frac{\langle -b_{\hat r}\,b_{\hat\varphi}\rangle_\rho}{\langle p_{\rm gas}+p_{\rm mag}\rangle_\rho},\\
\alpha_{\rm Rey} &= \frac{\langle (\rho+u_{\rm gas}+p_{\rm gas}+b^2)\,\delta v_{\hat r}\,\delta v_{\hat\varphi}\rangle_\rho}{\langle p_{\rm gas}+p_{\rm mag}\rangle_\rho},\\
\delta v_{\hat\imath}&=v_{\hat\imath}-\langle v_{\hat\imath}\rangle_\rho , \nonumber
\label{eq:alpha_components}
\end{align}
where  $p_{\rm mag}=b^2/2$. The Reynolds term is split into gas ($\rho+u_{\rm gas}+p_{\rm gas}$) and magnetic ($b^2$) contributions, $\alpha_{\rm Rey}=\alpha_{\rm Rey,gas}+\alpha_{\rm Rey,mag}$. Consistent with Equation~\ref{eq:maxwell_laminar}, the laminar Maxwell term is built from the time- and azimuth-averaged field, $\bar b_{\hat\imath}(r,\theta)$ (density weighted over $t$ and $\varphi$):
\begin{align}
\alpha_{\rm Max,lam} &= \frac{\langle -\bar b_{\hat r}\,\bar b_{\hat\varphi}\rangle^\theta_\rho}{\langle p_{\rm gas}+p_{\rm mag}\rangle_\rho},\\ \nonumber
\alpha_{\rm Max,turb}&=\alpha_{\rm Max}-\alpha_{\rm Max,lam}.
\label{eq:alpha_maxwell_components}
\end{align}
where $\langle\rangle^\theta_\rho$ means density weighted averaging in $\theta$ direction.
All quantities are averaged over $t=20{,}000$--$25{,}000\,r_g/c$.

\section{\label{sec:results}Quasi-Steady Angular Momentum Flux Transport Across Accretion States}

\subsection{Angular Momentum Flux Transport}
\label{sec:flux_transport}

\begin{figure*}[ht!]
\centering

\includegraphics[width=\textwidth]{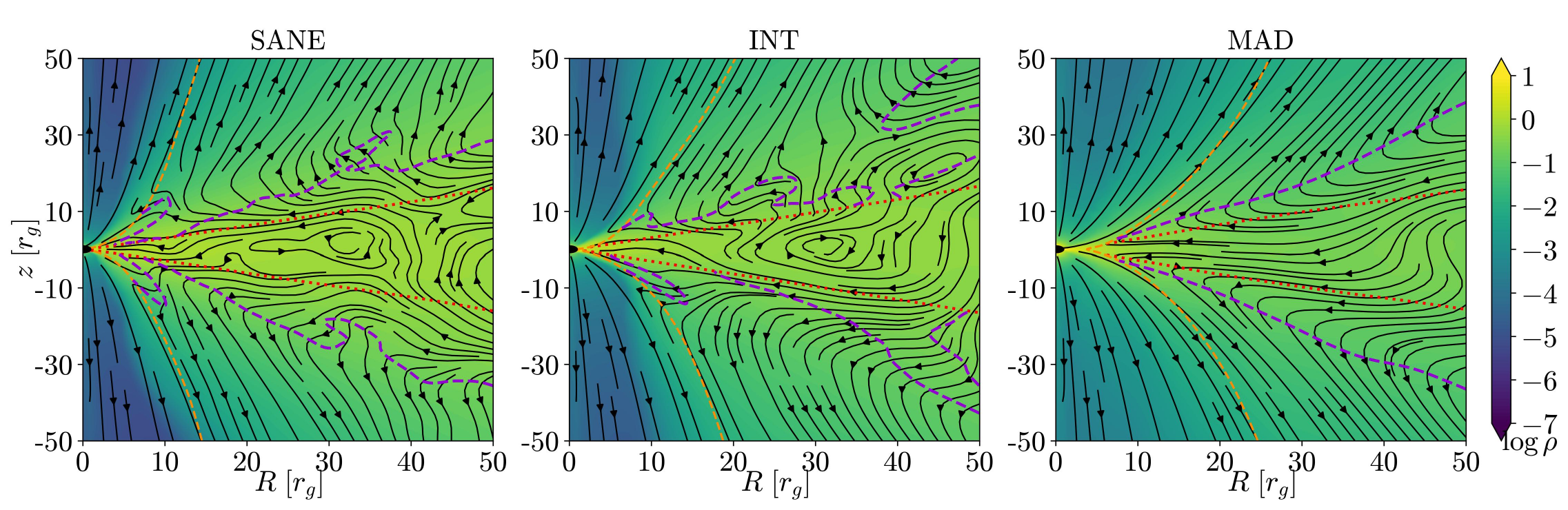}

\caption{
Time-averaged angular momentum flux streamlines for the three accretion configurations around a rapidly spinning black hole ($a=0.998$). The streamlines trace the flow of angular momentum with density contours (color scale) and region boundaries (orange: jet, red: disk, violet: wind). The MAD configuration shows a coherent polar transport channel, while the SANE configuration exhibits smaller-scale circulation in the disk and wind.}
\label{fig:angular_momentum_streamlines}
\end{figure*}

Figure~\ref{fig:angular_momentum_streamlines} presents the time-averaged streamlines of the total orthonormalized angular momentum flux ($\dot{\hat{\mathcal{J}}^i}=\dot{\mathcal{J}^i}\sqrt{g_{ii}}$) for our three configurations: SANE, INT, and MAD. 

Figure~\ref{fig:angular_momentum_streamlines} reveals distinct angular-momentum-flux morphologies. SANE exhibits small-scale circulation in the disk, INT contains several localized regions of inward flux, and MAD develops a coherent outward polar channel together with inward disk advection.

\begin{figure*}[ht!]
\centering

\includegraphics[width=\textwidth]{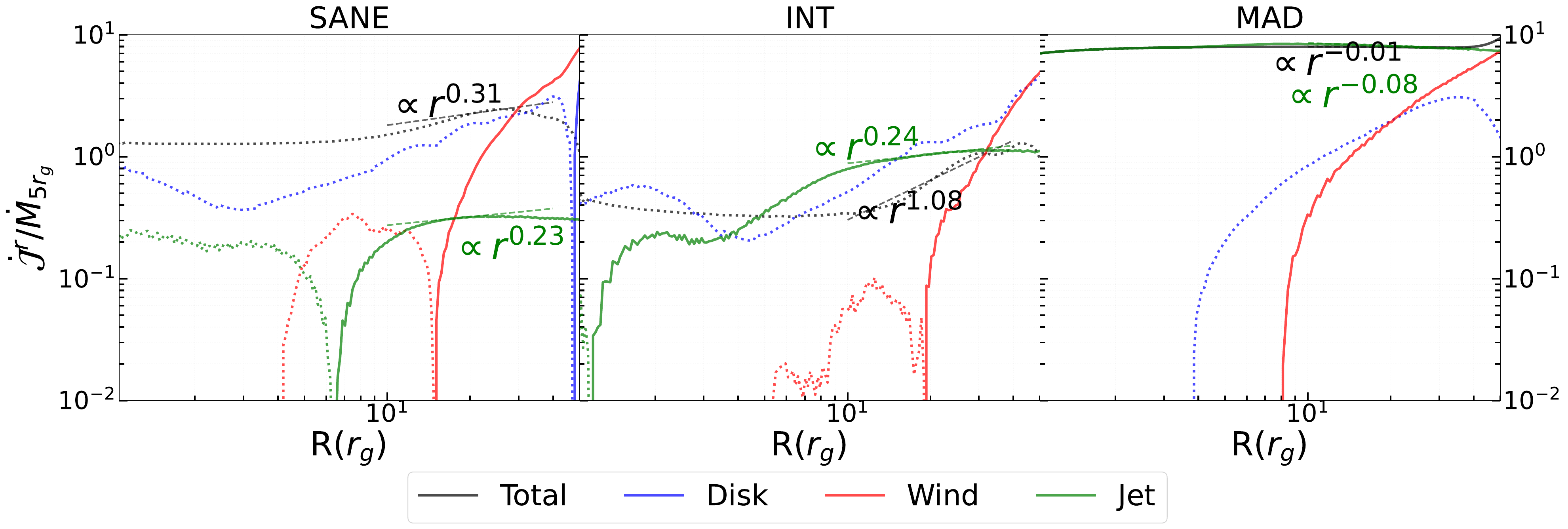}

\caption{Time-averaged radial profiles of specific angular momentum flux for 
SANE (left), INT (middle), and MAD (right) configurations. Each panel shows 
contributions from total flux (black), jet (green), wind (red), and disk (blue) 
regions; solid lines indicate positive (outward) flux, dotted lines indicate 
negative (inward) flux. Power-law slopes are calculated over $10$--$40\,r_g$ 
(or $20$--$40\,r_g$ for wind components starting at $r \sim 20\,r_g$). SANE displays weaker jets than MAD/INT, stronger disk inward 
transport than MAD/INT, and persistent disk inward flux throughout, 
representing net inward angular momentum transport.INT shows an inward total flux whose magnitude grows with radius, dominated by inward disk transport despite outflow presence. MAD exhibits nearly constant total flux dominated by the jet contribution (a jet-dominated angular-momentum budget), with wind beginning at $r \sim 10\,r_g$ and 
persistent disk inward transport.
}
\label{fig:jdot_profiles}
\end{figure*}

SANE remains disk dominated, with a net inward angular-momentum flux at all radii (Figure~\ref{fig:jdot_profiles}). Its wind changes sign at $\sim 15\,r_g$ and carries angular momentum outward at larger radii, but this contribution remains too small to overcome the inward disk flux. INT is transitional: its jet carries angular momentum outward over most radii, whereas the total flux remains inward because of the disk contribution. MAD is jet dominated, and its nearly flat radial profile is consistent 
with a coherent Blandford--Znajek electromagnetic flux extending from 
the horizon to large radii. We emphasize that this jet flux is sourced 
by the black-hole spin rather than by the accreting gas: a positive, 
jet-dominated total flux therefore does not by itself measure how 
efficiently angular momentum is removed from the disk. Disk transport 
is quantified separately by the stress components within the disk 
(Section~\ref{sec:components}) and by the wind and disk-surface fluxes 
(Section~\ref{sec:surface_flux}). Thus, the progression from SANE to MAD primarily changes the relative importance of inward matter advection, disk stresses, and the spin-powered jet contribution.

The fitted indices quantify these trends. Because $\partial_\varphi$ is a Killing vector, $\nabla_\mu T^\mu{}_\varphi=0$ has no source term, and integrating over a spherical shell gives $\partial_t L(r) = -\partial_r \dot{\mathcal{J}}^r(r)$, where $L(r)$ is the angular momentum per unit radius (neglecting density floors); the analogous relation holds for $\dot M$. A radially constant total flux therefore indicates inflow equilibrium, independent of how the flux is shared among jet, wind, and disk. In MAD the total flux is constant to within $1\%$ ($\propto r^{-0.01}$) over $10$--$40\,r_g$, so the flow is in inflow equilibrium over this range. In SANE and INT the magnitude of the inward total flux increases outward ($\propto r^{0.31}$ and $r^{1.08}$): more angular momentum enters the region $10$--$40\,r_g$ than leaves it, and $\sim27\%$ (SANE) and $\sim70\%$ (INT) of the inward flux at $40\,r_g$ is stored inside, so these flows have not yet reached inflow equilibrium over this range. The MAD jet flux is nearly constant ($\propto r^{-0.08}$) and carries $96$--$105\%$ of the total flux between $10$ and $40\,r_g$, as expected for an electromagnetic flux carried along field lines anchored on the horizon; its $\sim10\%$ decline across this range is consistent with the jet-to-wind transfer at the wind--jet interface (Section~\ref{sec:surface_flux}). In SANE and INT the jet flux instead grows outward ($\propto r^{0.23}$ and $r^{0.24}$), indicating that the jet gains angular momentum across its boundary. In general, the radial gradient of the flux in any region equals the net flux through its lateral boundaries (Appendix~\ref{app:surface_flux}).

Compared with the non-spinning ($a=0$) MAD and SANE systems of \citet{Chatterjee2022}, the key difference is the presence of an additional angular-momentum channel: the Blandford--Znajek torque extracts angular momentum from the black hole itself and carries it outward as a coherent electromagnetic flux through the jet. The eruption-driven transport identified by \citet{Chatterjee2022} operates in both cases---flux eruptions seed the outward turbulent stresses and power the winds that remove angular momentum from the disk---and this mechanism naturally produces a radially declining flux as the eruption-driven stresses deposit angular momentum into the wind over a range of radii. In our spinning MAD, this eruption-driven component is supplemented by the spin-powered jet flux, which suffers little radial dissipation and dominates the total budget, yielding the approximately constant radial profile in Figure~\ref{fig:jdot_profiles}. The two channels also differ in their source: the jet flux is drawn from the black-hole spin, whereas the eruption-driven wind and turbulent stresses redistribute the disk's own angular momentum. This difference is consistent with a spin-powered transport channel, although a matched spin sequence is required to isolate the effect of black-hole spin from differences in the simulation setups.

\subsection{Components of Angular Momentum Flux}\label{sec:components}

\begin{figure*}
\centering

\includegraphics[width=\textwidth]{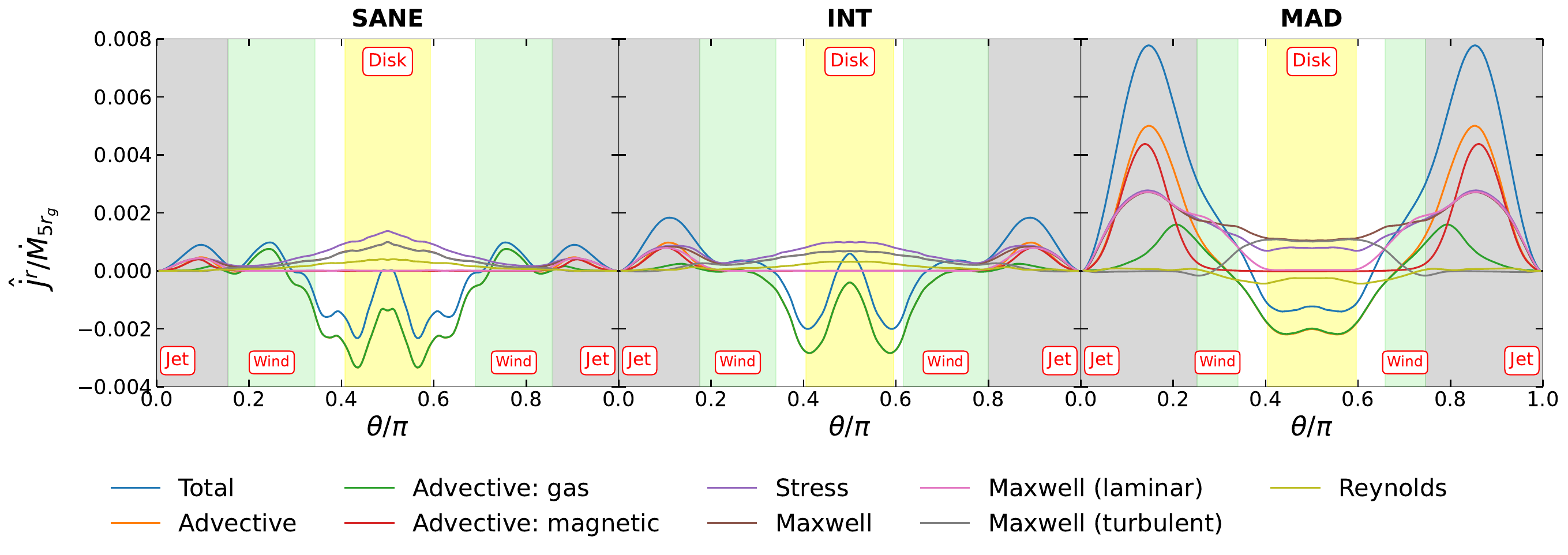}

\caption{Angular structure of angular momentum flux components at $r = 20\,r_g$ 
for (left) SANE, (middle) INT, and (right) MAD configurations. Shaded regions 
indicate jet (gray), wind (light green), and disk (yellow) boundaries. Solid 
lines denote positive (outward) flux; dotted lines denote negative (inward) flux. 
Components shown: total (blue), advective (orange), advective gas (green), 
advective magnetic (red), stress (purple), Maxwell (brown), Maxwell laminar (pink), 
Maxwell turbulent (gray), and Reynolds (olive). SANE: Reynolds stress small, predominantly positive, 
unable to compensate for dominant negative advection. INT: disk shows negative 
total flux dominated by inward gas advection, with positive stress insufficient 
to reverse the inward flow. Reynolds stress positive.  MAD: jets show strong positive 
total flux with significant advective-magnetic and Maxwell stress contributions; 
winds maintain positive flux with comparable advective (gas) and stress torques; 
disk exhibits near-zero fluxes with overall inward flux from gas advection (disk 
not primary transport pathway); Reynolds stress subdominant everywhere (negative in disk).}
\label{fig:jdot_components_polar}
\end{figure*}

\begin{figure*}
\centering

\includegraphics[width=\textwidth]{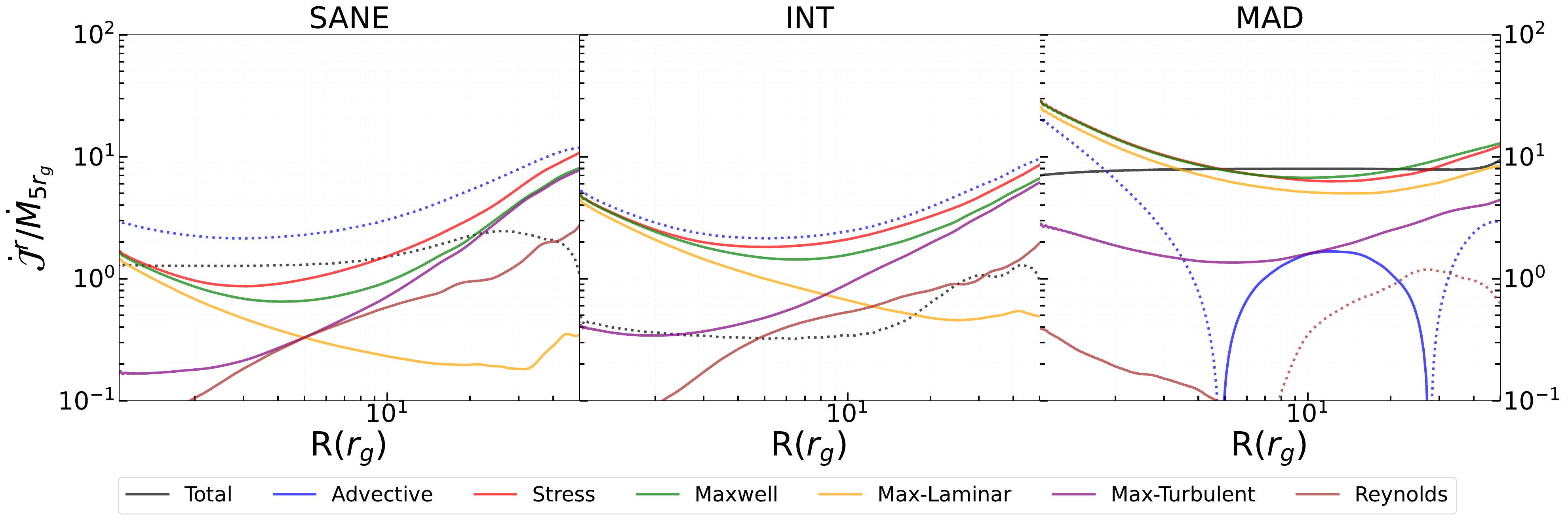}

\caption{Radial evolution of shell-integrated angular momentum flux components for (left) SANE, (middle) INT, and (right) MAD configurations. Solid lines indicate positive (outward) flux; dotted lines indicate negative (inward) flux. Components shown: Total (black), Advective (blue), Stress (red), Maxwell (green), Max-Laminar (orange), Max-Turbulent (purple), and Reynolds (brown). SANE: Exhibits persistent inward total flux dominated by strong inward advection that increases moderately with radius. INT: Shows persistent inward total flux with advective (inward) and stress (outward) components strengthening at similar rates. MAD: Maintains steady outward total flux with nearly constant stress; advective flux is highly variable, weakening at $10$--$20\,r_g$ and becoming positive due to magnetic advection.}
\label{fig:jdot_components_radial}
\end{figure*}

Figure~\ref{fig:jdot_components_polar} displays the angular structure of the components (discussed in Section~\ref{sec:diagnostics}) at $r = 20\,r_g$, while Figure~\ref{fig:jdot_components_radial} shows their time-averaged radial profiles. In Figure \ref{fig:jdot_components_radial}, solid lines indicate positive (outward) flux while dotted lines represent negative (inward) flux.

The angular structure at $r = 20\,r_g$ (Figure~\ref{fig:jdot_components_polar}) 
reveals fundamental distinctions between the three configurations. SANE presents 
advection-dominated transport: jets show very weak fluxes across all components, winds display large outward flux strongly dominated by gas 
advection with stress far too small in magnitude, and the disk exhibits large 
negative flux dominated by strong inward gas advection with deeply subdominant 
outward stress. INT displays balanced transport: jets have moderate 
positive flux with equal advective and stress contributions, winds are 
stress-dominated, and the disk exhibits negative flux dominated by inward gas 
advection with positive stress insufficient to reverse inward flow. MAD exhibits 
stress-dominated transport: jets show strong positive total flux with significant 
advective (magnetic) and Maxwell stress contributions, winds maintain positive flux 
with comparable advective (gas) and stress torques, and the disk carries outward Maxwell stress that is largely offset by inward gas advection, yielding a small net flux. Thus, while stress-driven transport is present in the disk, the net outward angular-momentum flux is dominated by the jet, with a secondary wind contribution. At $r=20\,r_g$, the Reynolds stress is generally subdominant (negative in the MAD disk and positive in the SANE and INT disks) and is typically $10$--$100\times$ smaller than the Maxwell stress in the strongly magnetized regions. This indicates that magnetic stresses dominate the outward stress contribution in these regions.

The radial evolution (Figure~\ref{fig:jdot_components_radial}) demonstrates 
contrasting transport mechanisms. SANE shows outward stress with increasing 
turbulent Maxwell and Reynolds components with radius; laminar Maxwell stress remains stronger 
at inner radii but decreases steadily while turbulent components dominate at outer 
radii, with declining stress efficiency combined with persistent advective dominance 
explaining why SANE cannot achieve net outward extraction despite generating 
MRI-driven turbulent transport. In INT, the laminar Maxwell stress starts moderately strong but decreases 
significantly while turbulent Maxwell stress grows and dominates at large radii, 
with this balanced growth preventing stress from overcoming advection despite 
partial flux saturation. In MAD, the torque due to laminar Maxwell 
stress dominates at small radii and remains substantial at large radii, with 
turbulent Maxwell stress growing to achieve near-equipartition at large radii, 
establishing a stress-dominated regime in which stress easily overcomes 
weak advection to produce a net outward flux. The laminar component in 
the polar region is fed by the Blandford--Znajek torque on the black 
hole, while the removal of the disk's own angular momentum proceeds 
through the disk stresses and the wind and disk-surface fluxes 
(Section~\ref{sec:surface_flux}). The disk turbulence governed by the turbulent component of Maxwell stress carries out more angular momentum in SANE, whereas the laminar winds in MAD are the major component of carrying out angular momentum in MAD which are seeded by flux eruptions. The contrasting radial evolution of the laminar Maxwell stress---approximately constant in MAD but declining in SANE---helps explain why the MAD run maintains a net outward flux. The result links magnetic-flux saturation and large-scale field organization to both the magnitude and radial persistence of the electromagnetic angular-momentum flux.

\subsection{Interaction of the Jet, Wind, and Disk}\label{sec:surface_flux}

The surface fluxes (Figure~\ref{fig:jdot_surfaces}; see Appendix~\ref{app:surface_flux} for the full component decomposition) show that magnetic-flux saturation controls not only the radial transport (Section~\ref{sec:components}) but also the vertical, inter-zone redistribution of angular momentum. The dominant stress at the jet surface progresses systematically with flux saturation: Reynolds-driven in SANE, mixed turbulent-Maxwell and Reynolds in INT, and comparable laminar and turbulent Maxwell contributions in MAD, reflecting the growing role of organized large-scale field.
The wind-jet interface provides the clearest diagnostic of magnetic braking. In MAD, the large-scale field threading both regions transfers angular momentum from jet to wind through laminar Maxwell stress; in SANE this coupling is essentially absent. At the wind - disk edge, all three configurations move angular momentum from the disk edge into the wind, but by different means: advection in SANE and INT (with the stress opposing), and organized laminar Maxwell stress in MAD, where, notably, the turbulent Maxwell component partially counteracts the laminar extraction, revealing an internal competition between coherent and fluctuating fields. At the disk surface, only MAD achieves net stress-driven extraction through its upper and lower boundaries, establishing a vertical angular-momentum sink that reinforces the outward radial transport of Section~\ref{sec:components}; the Reynolds stress is negative at this surface in all three runs, consistent with turbulent mixing opposing magnetic extraction.
Taken together, the MAD run exhibits a coherent multi-zone angular-momentum network: organized fields transfer angular momentum between the jet and wind and extract it vertically through the disk surface. In SANE, the weaker large-scale field substantially reduces this coupling, leaving advection to dominate the inter-zone exchange.

\subsection{Viscous Transport and Accretion Efficiency}
\label{sec:alpha_viscosity}

\begin{figure*}[htbp]
\centering

\includegraphics[width=\textwidth]{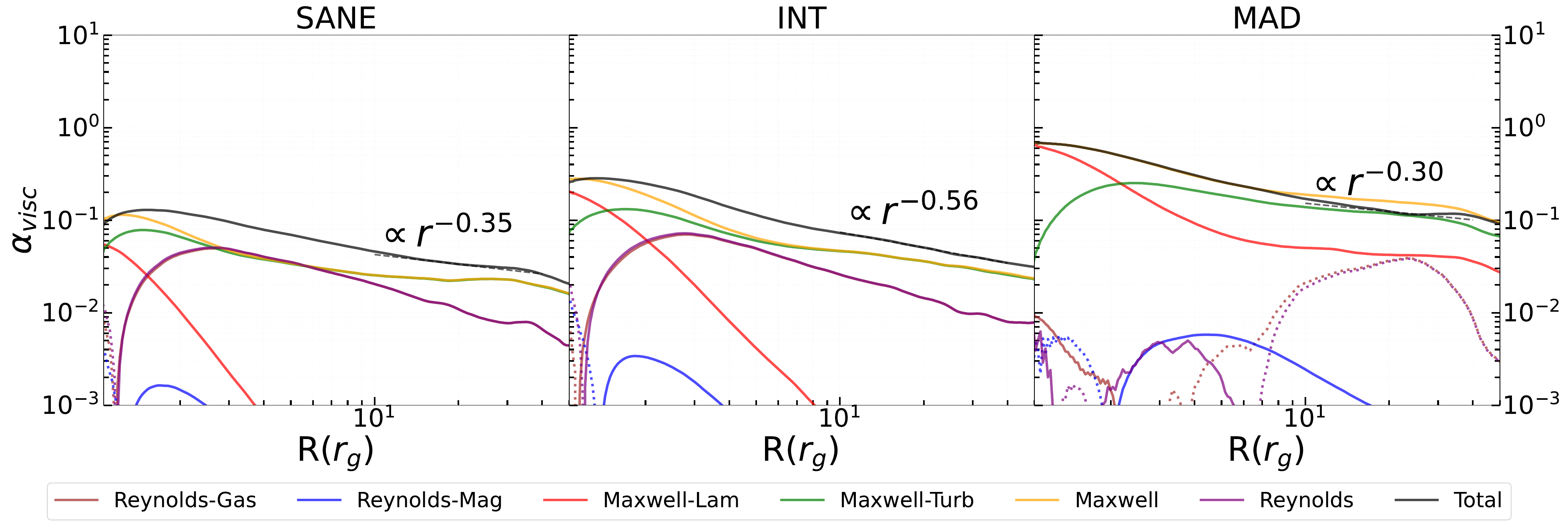}

\caption{Radial profiles of $\alpha_{\rm visc}$ and its components for the three configurations. Shown are the Maxwell term (orange), split into laminar (red; from the time- and azimuth-averaged field) and turbulent (green) parts, the Reynolds term (purple), split into gas (brown) and magnetic (blue) parts, and the total (black). Solid lines indicate positive (outward) transport; dotted lines indicate negative (inward) transport. The dashed line shows a power-law fit to the total $\alpha_{\rm visc}$ over $10$--$40\,r_g$. SANE and INT Maxwell dominated in the very inner region, with the laminar part playing a role that dies out in the bulk disk region. The turbulent component of Maxwell stress plays a significant role throughout, but Reynolds stress also becomes significant in the bulk disk. MAD shows domination by laminar Maxwell stress in the very inner region (possibly due to flux eruptions); the turbulent Maxwell dominates in the outer region, and the Reynolds term is negligible and mostly negative in the bulk disk.}
\label{fig:alpha_viscosity}
\end{figure*}

The radial evolution of the effective stress parameter $\alpha_{\mathrm{visc}}$ illustrates a fundamental transition in the mechanism of angular momentum transport as magnetic flux accumulates in the disk. In the weakly magnetized (SANE) and intermediate (INT) regimes, stress in the bulk disk ($r \gtrsim 3\,r_g$) is dominated by fully developed, isotropic magnetorotational turbulence (shown by dominant Maxwell turbulent component), where Maxwell stresses are nearly entirely stochastic and Reynolds stresses contribute significantly via non-axisymmetric velocity fluctuations. As the system transitions toward the MAD state, the nature of stress transport changes qualitatively: the strength of turbulent fluctuations diminishes relative to a large-scale, coherent magnetic field that maintains a substantial laminar Maxwell stress out to large radii ($r \sim 40\,r_g$) ($\sim27$--$30\%$ of the Maxwell stress over $10$--$40\,r_g$, compared with $\lesssim3\%$ in SANE and INT). Crucially, the strong magnetic obstruction in the MAD disk suppresses hydrodynamical velocity fluctuations to the point where the Reynolds stress changes sign in the main disk body ($r \sim 7\text{--}70\,r_g$), physically representing an inward hydrodynamic transport of angular momentum against the outward magnetic torque. Near the event horizon ($r \lesssim 2\,r_g$), plunging dynamics and strong frame-dragging force Maxwell stresses to dominate across all models, though the transport becomes progressively non-turbulent—shifting from a balance of turbulent and ordered field structures in SANE to an almost purely laminar, force-free flux tube in MAD. Thus, while SANE and INT disks transport angular momentum through classic local turbulent viscosity, MAD disks operate via global, large-scale magnetic torques. 

Because $\alpha_{\rm visc}$ is density weighted, it samples the disk body, whereas the shell-integrated fluxes of Figure~\ref{fig:jdot_components_radial} are dominated within $\sim10\,r_g$ by the magnetized funnel; the two diagnostics therefore weight the laminar and Reynolds contributions differently. Since the flow inside a few $r_g$ is plunging, the near-horizon values of $\alpha_{\rm visc}$ measure magnetic stress relative to pressure rather than a diffusive viscosity in the sense of \citet{Shakura1973}; we therefore restrict the viscous-timescale estimates below to $r=10$--$40\,r_g$.

\begin{figure*}
    \centering

    \includegraphics[width=\textwidth]{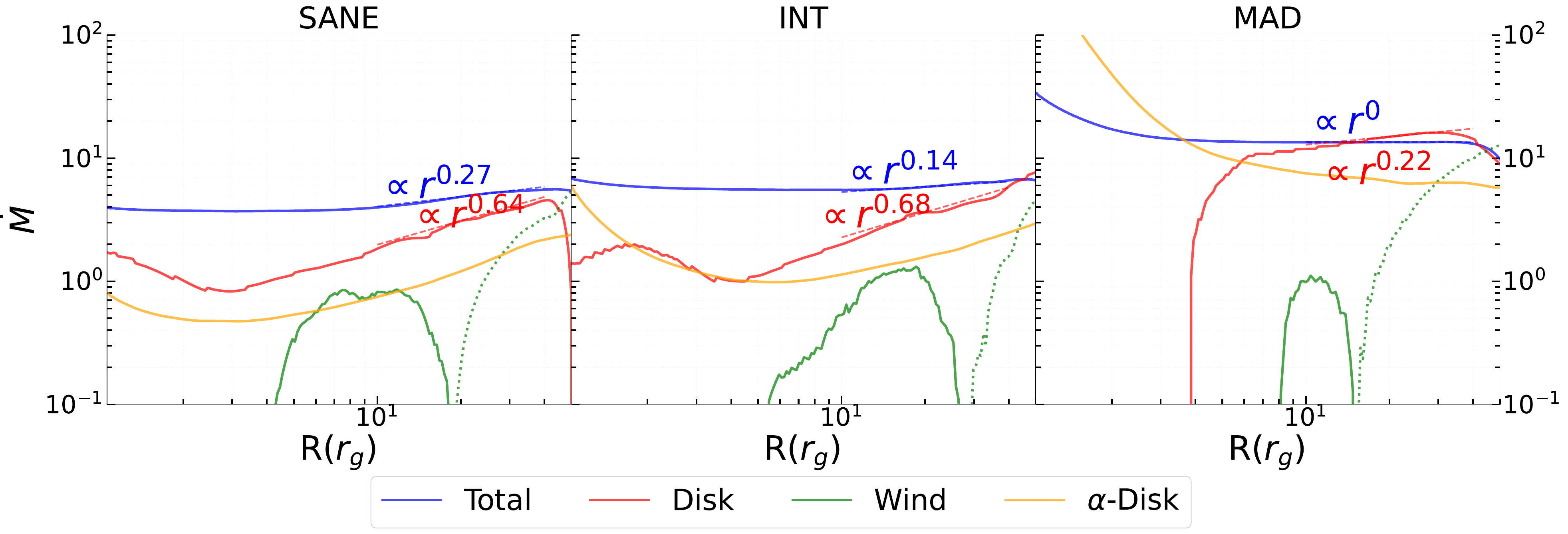}

    \caption{Radial profiles of mass accretion rates for SANE (left), INT (middle), and MAD (right). Each panel shows the total (blue), disk (red), and wind (green) accretion rates and the $\alpha$-disk estimate (orange), computed with the total $\alpha_{\rm visc}$ of this section. Solid lines indicate positive (inward) flux and dotted lines negative (outward) flux. Dashed lines show power-law fits to the total and disk accretion rates over $10$--$40\,r_g$. The total accretion rate is radially constant in MAD, indicating inflow equilibrium, whereas it increases outward in SANE and INT, where mass is still accumulating in the inner flow. The disk accretion rate increases outward in all runs, most steeply in SANE and INT, as an increasing fraction of the inflow is diverted out of the disk.}
    \label{fig:mdot_comparison}
\end{figure*}

To assess how efficiently our simulations drive mass inflow, we compute the mass accretion rate within the disk ($\dot{M}_{\rm disk}$), winds ($\dot{M}_{\rm wind}$), and the total system ($\dot{M}_{\rm tot}$), comparing them with the classical $\alpha$-disk prediction \citep{Shakura1973, Pringle1981}. Figure~\ref{fig:mdot_comparison} presents the radial profiles of mass accretion rates for our three configurations alongside the $\alpha$-disk model calculated using $\dot{M}_{\rm SS} = 4\pi \alpha_{\rm visc} H (P_{\rm gas}+P_{\rm mag})/\Omega$, using the same pressure that normalizes $\alpha_{\rm visc}$, where we employ the total $\alpha_{\rm visc}$ values computed in this section. By the conservation argument of Section~\ref{sec:flux_transport}, a radially constant $\dot M_{\rm tot}$ indicates inflow equilibrium. Over $10$--$40\,r_g$ this holds in MAD ($\dot M_{\rm tot}\propto r^{0.00}$, constant to within $3\%$), whereas $\dot M_{\rm tot}\propto r^{0.27}$ (SANE) and $r^{0.14}$ (INT): $\sim28\%$ (SANE) and $\sim16\%$ (INT) of the inflow at $40\,r_g$ is stored between $10$ and $40\,r_g$ instead of reaching smaller radii, consistent with the angular-momentum accumulation found in Section~\ref{sec:flux_transport}. The disk accretion rate scales as $\dot M_{\rm disk}\propto r^{s}$ with $s=0.64$ (SANE), $0.68$ (INT), and $0.22$ (MAD). An outward increase of the inflow rate is the signature of mass loss from the disk at each radius \citep{Blandford1999, Yuan2012}; the smaller $s$ in MAD indicates that its disk loses a smaller fraction of its inflow between $40$ and $10\,r_g$. The MAD configuration has the largest $\alpha_{\rm visc}$ at all radii, reflecting sustained stress generation by organized magnetic fields. The wind mass loading, $-\dot M_{\rm wind}/\dot M_{\rm disk}$, increases with radius in all three states. It is $0.08$, $0.51$, and $0.72$ at $20$, $30$, and $40\,r_g$ in SANE, and $0.13$, $0.40$, and $0.69$ in MAD. In INT the wind region still carries a net inflow at $20\,r_g$ ($-\dot M_{\rm wind}/\dot M_{\rm disk}=-0.28$), with outflow loadings of $0.04$ and $0.27$ at $30$ and $40\,r_g$. The SANE wind therefore develops at larger radii than the MAD wind but reaches a comparable loading by $40\,r_g$, despite its weaker large-scale field. This trend may reflect different wind-driving mechanisms: the SANE outflow is broader and more diffuse at large radii, whereas the MAD wind remains more closely associated with organized large-scale fields. A dedicated force or Bernoulli analysis is required to distinguish turbulent, thermal, and magnetocentrifugal driving. The wind mass-loading thus emerges as a critical parameter distinguishing the angular momentum budgets of turbulence-dominated versus magnetically-dominated accretion states.

In the bulk disk, the effective $\alpha$ parameters have direct 
implications for disk evolution timescales. Over the bulk disk the total stress-to-pressure ratio declines as $\alpha_{\rm visc}\propto r^{-0.35}$ (SANE), $r^{-0.56}$ (INT), and $r^{-0.30}$ (MAD) (Figure~\ref{fig:alpha_viscosity}). Since $t_{\rm visc}/t_{\rm dyn}\simeq\alpha_{\rm visc}^{-1}(r/H)^2$, these indices set how the viscous timescale grows with radius relative to the orbital time.The viscous timescale, $t_{\rm visc} \sim \alpha_{\rm visc}^{-1}(r/H)^2\,t_{\rm dyn}$ with $t_{\rm dyn} = \Omega_K^{-1} = (r^{3/2}+a)\,r_g/c$, evaluated locally 
over the quasi-steady region $r = 10$--$40\,r_g$, gives 
$t_{\rm visc}/t_{\rm dyn} \approx 84$--$93$ (MAD), $150$--$300$ (INT), 
and $290$--$440$ (SANE). For a $10\,M_\odot$ black hole these correspond 
to $t_{\rm visc} \sim 0.15$--$1.2$~s (MAD), $0.25$--$4$~s (INT), and 
$0.5$--$5$~s (SANE); for a $10^6\,M_\odot$ black hole, to 
$\sim 4$~hours--$1.3$~days (MAD), $7$~hours--$4$~days (INT), and 
$13$~hours--$6$~days (SANE). We note that the MAD flow is significantly 
sub-Keplerian ($\Omega/\Omega_K \approx 0.65$ over this range) owing to 
magnetic support, so these values are lower limits at the $\sim$50\% 
level for MAD; SANE and INT are nearly Keplerian. Extrapolating the 
viscous scaling to $r \sim 10^2$--$10^3\,r_g$ yields months-to-years 
timescales for supermassive black holes, broadly consistent with the 
spectral transitions observed in changing-look AGN 
\citep{LaMassa2015, Lawrence2018}. The systematically faster viscous 
evolution in MAD---shorter than SANE by a factor of $\sim$3.5--4.5 at fixed 
radius---implies that magnetically arrested systems should exhibit more 
rapid secular accretion variability and shorter reaccumulation timescales 
\citep{King1998}.

Our results for spinning black holes show a higher $\alpha$ value compared to the non-spinning MAD case studied by \citet{Chatterjee2022}. Additionally, \citet{Chatterjee2022} observed negative Reynolds stress throughout their non-spinning MAD, whereas our spinning MAD shows positive Reynolds stress out to $r \sim 11\,r_g$. This may indicate that black-hole spin modifies the radius at which interchange motions become more important than MRI-driven turbulence.

\section{Dynamical Evolution of Angular Momentum Transport and Jet Morphology During Flux Eruption Events}
\label{sec:flux_eruption}

The magnetically arrested disk state is characterized by episodic flux eruption events, 
during which accumulated magnetic flux at the horizon is expelled outward through 
magnetic interchange motions, potentially accompanied by reconnection \citep{Igumenshchev2008,McKinney2012}. These events 
temporarily reduce the magnetic flux threading the black hole and can significantly 
alter the accretion dynamics and jet properties. In this section, we analyze the 
temporal variability of the MAD simulation, focusing on the relationship between 
flux eruptions, mass accretion rate, angular momentum transport through the jet 
surface, and jet morphological parameters. 

\subsection{Angular Momentum Flux Through the Jet Surface}
\label{sec:jdot_definition}

\begin{figure*}[htbp]
\centering

\includegraphics[width=\textwidth]{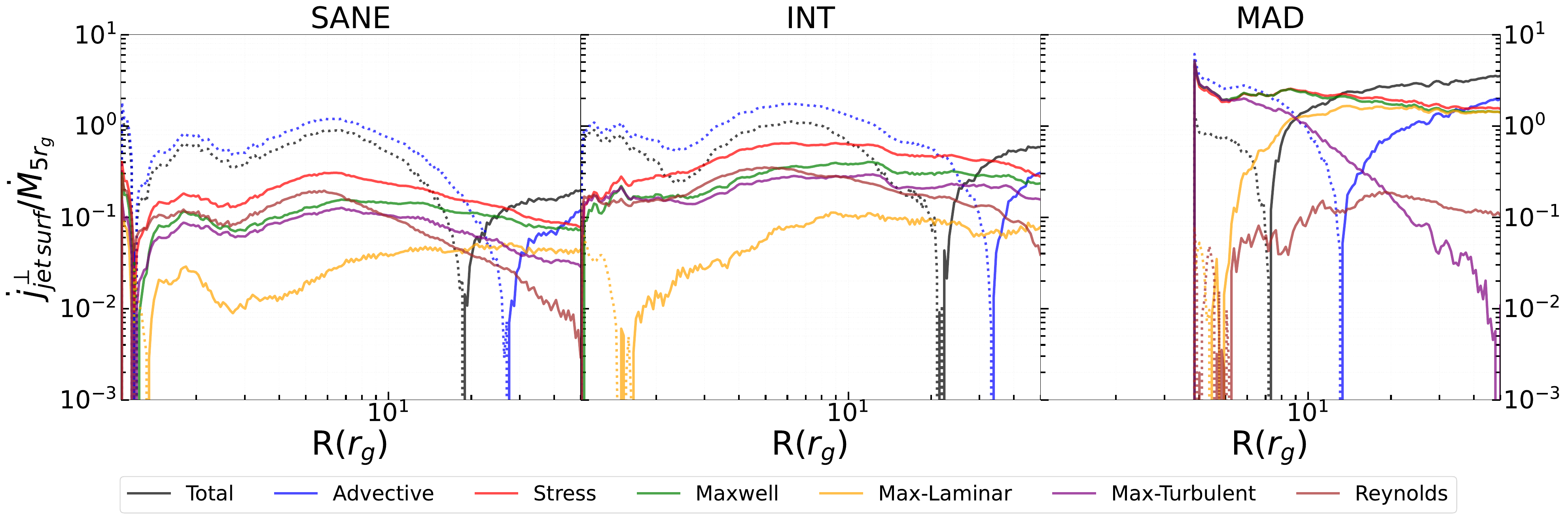}

\caption{Radial profiles of the non-cumulative angular momentum flux through 
the jet surface, $\dot{j}^\perp_{\rm jet\,surf}/\dot{M}_{5r_g}$, for the three 
magnetic field configurations: (left) SANE, (middle) INT, and (right) MAD. Each panel shows 
the total flux (black) and its decomposition into advective (blue), Maxwell 
(green), and Reynolds (brown) components. Solid lines indicate positive 
(outward) flux; dashed lines indicate negative (inward) flux plotted as 
absolute values. The MAD state exhibits net outward angular momentum transport 
for $r \gtrsim 7\,r_g$, driven by the dominant laminar Maxwell stress, while 
INT and SANE states show predominantly inward or negligible net transport.}
\label{fig:jdot_jetsurf_profiles}
\end{figure*}

Figure~\ref{fig:jdot_jetsurf_profiles} confirms the angular momentum extraction hierarchy established in Section~\ref{sec:surface_flux}, with MAD achieving net outward transport via dominant laminar Maxwell stress, while INT and SANE remain predominantly inward or near-zero. This outward MAD flux is temporarily suppressed, and briefly reversed, during the eruption events analyzed below.

\subsection{Identification of Flux Eruption Events}
\label{sec:eruption_id}

\begin{figure*}[htbp]
\centering

\includegraphics[width=\textwidth,height=18cm,keepaspectratio=false]{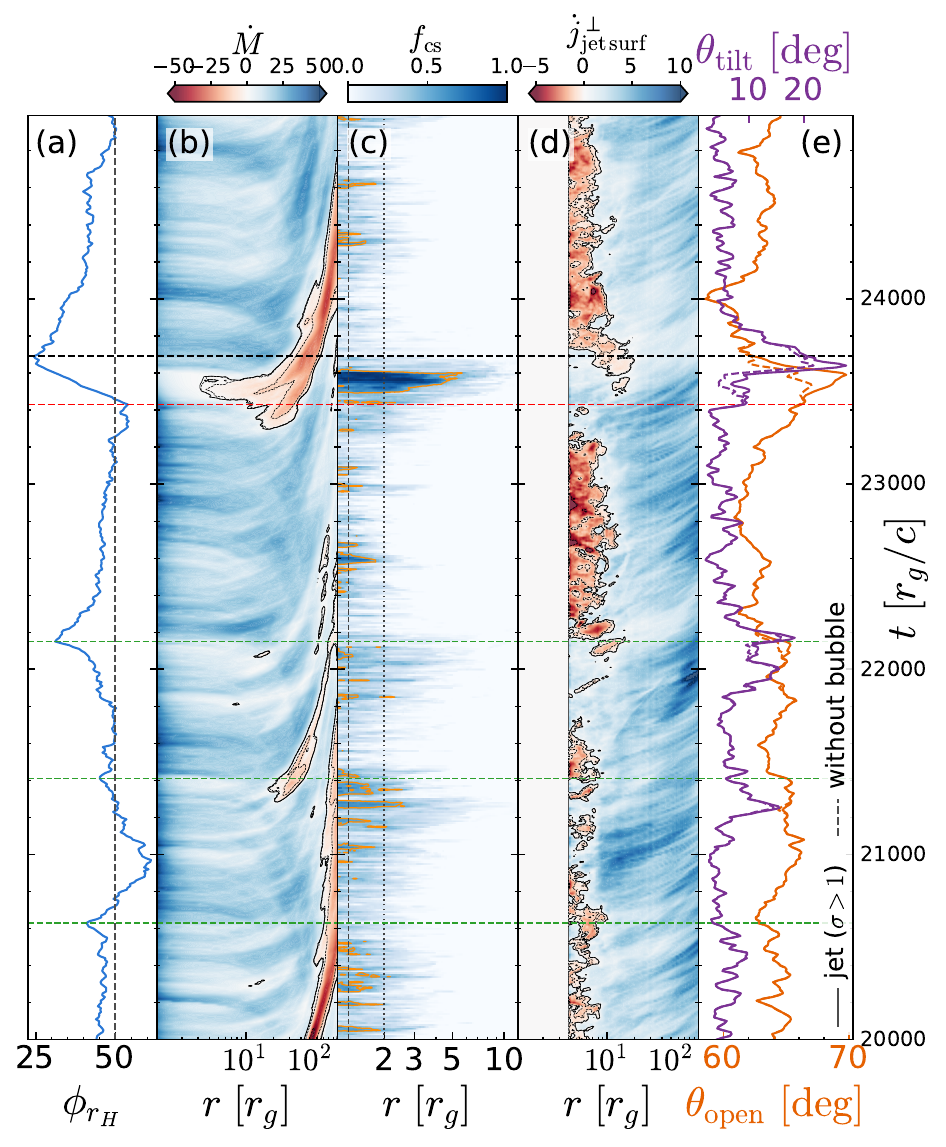}

\caption{Overview of flux eruption dynamics in the MAD simulation.
(a) Time series of the magnetic flux at the horizon $\phi_{r_H}$, with the vertical dashed line indicating the MAD saturation value $\phi_{\rm MAD} \approx 50$.
(b) Radius-time contour of the mass flux $\dot{M}(r,t)$.
(c) Radius-time diagram of the equatorial current sheet: the fraction of the azimuth $f_{\rm cs}$ in which the radial magnetic field reverses across the midplane in magnetically dominated gas ($\sigma > 1$; Equation~\ref{eq:cs_reversal}). The orange contour marks $f_{\rm cs} = 0.5$; the vertical dashed and dotted lines mark the ISCO ($1.24\,r_g$) and the equatorial ergosphere ($2\,r_g$), and the left edge of the panel is the horizon.
(d) Radius-time contour of the non-cumulative angular momentum flux through the jet surface, $\dot{j}^\perp_{\rm jet\,surf}/\dot{M}_{\rm avg}$.
(e) Jet opening angle $\theta_{\rm open}$ (orange, bottom axis) and tilt angle $\theta_{\rm tilt}$ (purple, top axis) for the jet ($\sigma > 1$; solid) and for the jet without the erupted bubble ($\sigma > 1$ and $\delta_t \ge 0.01$; dashed), averaged over $r_H < r \le 10\,r_g$ (Appendix~\ref{app:jet_morphology_def}).
In all panels, horizontal dashed lines mark the flux minima of the four identified eruption events (main event at $t \approx 23{,}690~r_g/c$ in black, secondary events in green), and the red dashed line marks the onset of the main event, $t_{\rm on} = 23{,}430~r_g/c$ (Section~\ref{sec:eruption_id}).}
\label{fig:flux_eruption_overview}
\end{figure*}

We identify flux eruption events using three distinct time windows, 
defined as follows. (i)~\textit{Detection window}: an event is the 
contiguous interval during which the horizon magnetic flux $\phi_{r_H}$ 
remains more than $1\sigma$ below its local running mean, where both the 
running mean and the local standard deviation are computed over a 
centered $500~r_g/c$ window; the event \textit{duration} is the length 
of this interval, and the \textit{detection depth} is the maximum 
fractional departure of $\phi_{r_H}$ below the running mean. 
(ii)~\textit{Baseline (comparison) window}: for the response analysis of 
Section~\ref{sec:event_response}, the pre-event reference is taken at 
the onset time $t_{\rm on}$, defined as the last time $\phi_{r_H}$ 
exceeded its 85th percentile before the eruption. 
(iii)~\textit{Minimum (recovery) window}: post-eruption values are 
averaged over $\pm 50~r_g/c$ around the flux minimum. Using the 
detection criterion, we find four distinct flux eruption events during 
$t = 20{,}000$--$25{,}000~r_g/c$ 
(Figure~\ref{fig:flux_eruption_overview}). The most prominent event has 
onset $t_{\rm on} \approx 23{,}430~r_g/c$, reaches its flux minimum at 
$t \approx 23{,}690~r_g/c$, and has a duration of $70~r_g/c$ with a 
detection depth of 27.3\%. Additional events have flux minima around 
$t \approx 22{,}150~r_g/c$ (22.3\% depth, $90~r_g/c$ duration), 
$t \approx 20{,}630~r_g/c$ (14.4\% depth, $40~r_g/c$ duration), and 
$t \approx 21{,}410~r_g/c$ (7.6\% depth, $50~r_g/c$ duration). We note 
that the interval from onset to flux minimum ($\approx 260~r_g/c$ for 
the main event) exceeds the detection-window duration because the flux 
declines gradually from its pre-eruption maximum before crossing the 
$1\sigma$ threshold.

The saturated horizon flux depends non-monotonically on spin and is similar at $a = 0$ and $a \approx 1$ \citep{Narayan2022, Zhang2024}, and our time-averaged $\langle\phi_{r_H}\rangle \approx 45$ over $t = 20{,}000$--$25{,}000~r_g/c$ is comparable. The eruptions in our run and in the non-spinning MAD of \citet{Chatterjee2022} therefore occur at a similar flux level. The fractional variability of the horizon flux, however, increases with $|a|$, from $\sigma_\phi/\mu_\phi = 0.094$ at $a = 0$ to $0.209$ at $a = 0.9$ \citep{Narayan2022}. We find $\sigma_\phi/\mu_\phi = 0.150$, larger than in the non-spinning case. Our interval is ten times shorter than the $50{,}000\,r_g/c$ used by \citet{Narayan2022} and contains only four eruptions, so this value depends on the averaging window and indicates only qualitatively that eruptions produce larger fractional flux drops around a rapidly spinning black hole.

During an eruption, low-density, magnetically dominated material expelled from the horizon can enter the instantaneous $\sigma > 1$ region and would then be counted as part of the jet. To identify this material, we compare the density in each cell with its time- and azimuthally averaged value at the same $(r,\theta)$,
\begin{equation}
\delta_t \equiv \frac{\rho}{\langle\rho\rangle_{t,\varphi}},
\label{eq:delta_t}
\end{equation}
where $\langle\rho\rangle_{t,\varphi}$ is averaged over $t = 20{,}000$--$25{,}000~r_g/c$. We define an erupted \textit{magnetized bubble} as the set of cells with $\sigma > 1$ and $\delta_t < 0.01$, i.e.\ magnetically dominated gas at least a hundred times less dense than is usual at that location. We use a density criterion alone because erupted flux tubes are strongly evacuated \citep{Ripperda2022, Loules2026}. The inclination of the magnetic field marks the boundary of the cavity rather than its interior, and a comparison with the instantaneous azimuthal average fails when the funnel itself is displaced to one side, as it is during eruptions. Outside eruptions (more than $300~r_g/c$ from any flux minimum) the criterion flags only $0.18\%$ of the magnetization-weighted jet volume within $r_H < r \le 10\,r_g$, whereas during the main eruption this share rises to $31\%$ at $t = 23{,}640~r_g/c$. The threshold $\delta_t = 0.01$ is the largest value that remains clean outside eruptions: thresholds of $0.03$ and $0.1$ already flag $5\%$ and $21\%$ of the jet in quiet periods, i.e.\ they remove ordinary jet material. Inside the polar funnel the density is set by the numerical floors, which cap the magnetization at $\sigma_{\max} = 25$, so erupted flux that enters the funnel itself cannot be distinguished by its density. The bubble share is therefore a lower limit.

\begin{figure*}[htbp]
\centering

\includegraphics[width=\textwidth]{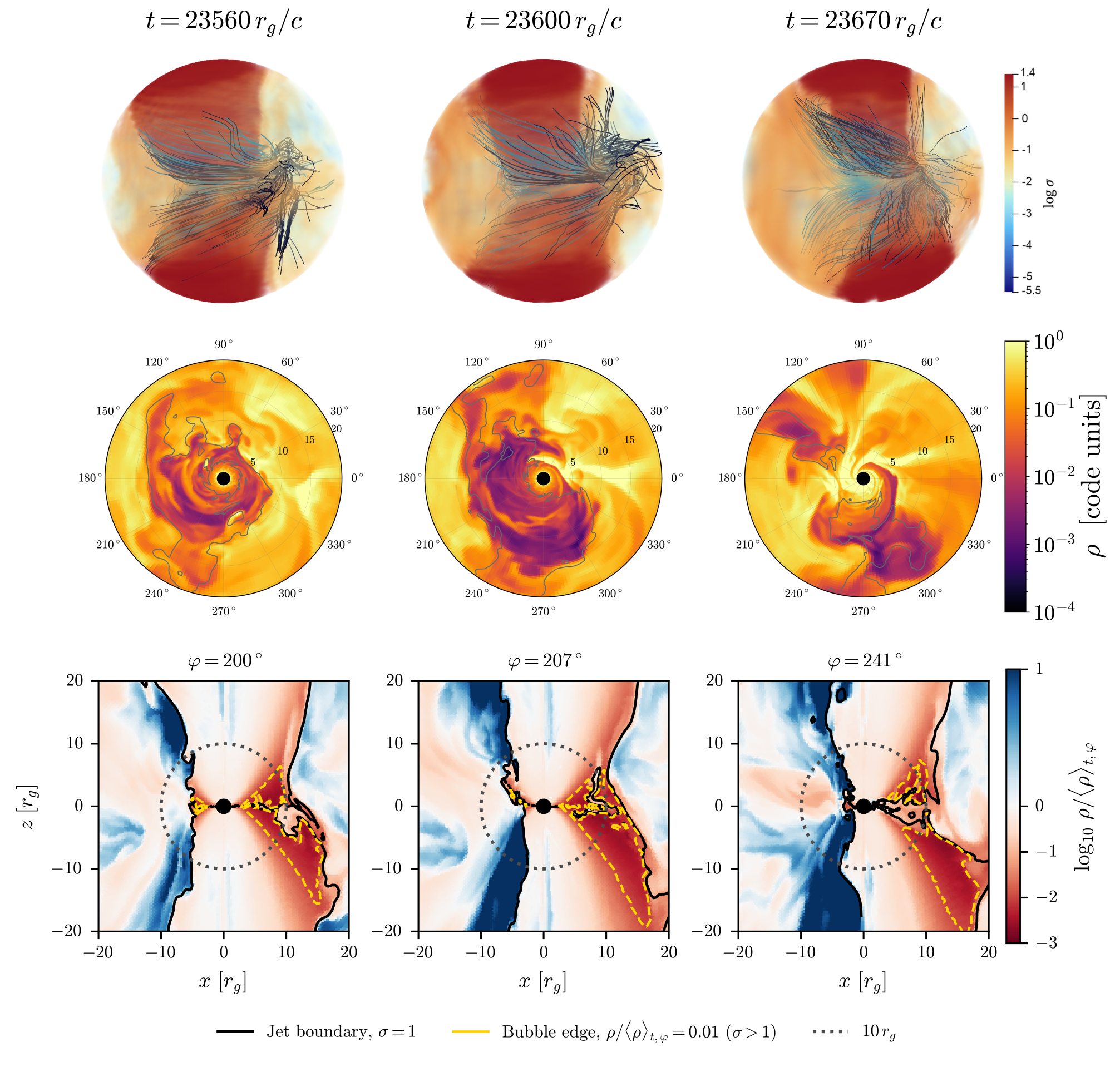}

\caption{Three-dimensional visualization and equatorial structure of magnetic 
flux eruption dynamics during the main event 
($t = 23{,}560$--$23{,}670~r_g/c$).
\textit{Top row}: 3D snapshots showing the progressive detachment and expulsion 
of a magnetic flux bundle (blue field lines) from the black hole-anchored 
magnetic field structure (black field lines). The color scale represents the 
magnetization parameter $\sigma$ in logarithmic scale. Left: The erupting flux 
bundle remains partially connected to the horizon-threading field. Center: The 
flux bundle has detached and is being expelled outward, disrupting the organized 
poloidal field geometry. Right: Complete eruption has occurred, leaving a 
reconfigured magnetic field structure. The jet (high-$\sigma$ polar regions) 
tilts substantially, and its magnetization-weighted opening angle decreases as the $\sigma > 1$ funnel contracts following the expulsion of horizon flux; the visually apparent broadening of the polar region reflects the growing tilt and asymmetry of the disrupted funnel rather than an increase in the measured width (Table~\ref{tab:event_statistics}).
\textit{Middle row}: Equatorial plane ($\theta = \pi/2$) rest-mass density 
$\rho$ (log scale) in polar coordinates, shown at 
the same three snapshots. Radial labels (in units of $r_g$) are indicated. 
The dark gray contour marks the isosurface of outflowing material at 
$v^r = 0.1c$ (in Kerr-Schild coordinates), tracing the leading edge of the 
ejected low-density flux bundle. Between $t = 23{,}560$ and $t = 23{,}670~r_g/c$, this contour advances from $r \sim 10\,r_g$ to $r \sim 20\,r_g$, corresponding to an apparent radial pattern speed of $\sim 0.1c$ in Kerr--Schild coordinates. A local orthonormal velocity or tracer analysis is required to infer the three-dimensional speed of the ejected material.
\textit{Bottom row}: Meridional slices of $\log_{10}\delta_t = \log_{10}(\rho/\langle\rho\rangle_{t,\varphi})$ (Equation~\ref{eq:delta_t}) at the same three snapshots; red marks gas that is less dense than usual at that location. The right half of each panel is the slice at azimuth $\varphi_0 = 200^\circ$, $207^\circ$ and $241^\circ$ (left to right), the $\sigma\sqrt{-g}$-weighted mean azimuth of the bubble cells within $20\,r_g$, and the left half is the slice at $\varphi_0 + 180^\circ$. Black contours mark the jet boundary ($\sigma = 1$), yellow contours the bubble edge ($\delta_t = 0.01$ inside $\sigma > 1$), and the dotted circle $r = 10\,r_g$, the outer radius over which the jet angles are averaged. The black disk is the black hole.}
\label{fig:eruption_3d}
\end{figure*}

Figure~\ref{fig:eruption_3d} visualizes the three-dimensional structure and 
temporal evolution of the main flux eruption event from two complementary 
perspectives. The top row shows the 3D magnetic field morphology: the 
progressive detachment of the erupting flux bundle (white field lines) from 
the horizon-anchored organized field (blue field lines) directly disrupts 
the poloidal geometry that sustains jet collimation, providing a physical 
explanation for the coordinated suppression of accretion, angular momentum 
transport, and jet stability observed during these events. The middle row 
presents the corresponding equatorial ($\theta = \pi/2$) density maps in 
polar projection, revealing the complementary disk-plane view of the same 
eruption. The low-density cavity associated with the expelled flux bundle is 
clearly visible sweeping outward through the disk, delineated by the 
$v^r = 0.1c$ outflow contour (gray lines). Between $t = 23{,}560$ and $t = 23{,}670\,r_g/c$, the outward-moving front advances from $r \sim 10\,r_g$ to $r \sim 20\,r_g$ over $\Delta t \approx 110\,r_g/c$, corresponding to an apparent radial pattern speed of $\sim 0.1c$ in Kerr-Schild coordinates. This estimate characterizes the motion of the selected contour rather than the local three-dimensional fluid velocity. The density maps also reveal the 
non-axisymmetric, spiral-arm morphology of the expelled material — a 
characteristic feature of magnetic interchange instabilities — which is associated with the azimuthal asymmetry and jet tilt seen in the 3D panels. The bottom row of Figure~\ref{fig:eruption_3d} shows $\log_{10}\delta_t$ in meridional slices through the bubble, with the $\sigma = 1$ boundary and the bubble edge ($\delta_t = 0.01$ inside $\sigma > 1$) overplotted. The bubble forms at the base of the magnetically dominated region close to the equatorial plane, where the bubble makes up $45$--$64\%$ of the $\sigma > 1$ volume between $3$ and $20\,r_g$ within $30^\circ$ of the midplane in the three snapshots, and it spreads into the lower hemisphere with time: its share of the $\sigma > 1$ volume at $\theta > 120^\circ$ grows from $9\%$ to $23\%$ between $t = 23{,}560$ and $23{,}670~r_g/c$. The upper funnel ($\theta < 60^\circ$) is only $3$--$4$ times less dense than usual (median $\delta_t = 0.24$--$0.31$), and the bubble makes up at most $2\%$ of its $\sigma > 1$ volume. The asymmetry of the upper jet in these snapshots is therefore mainly a displacement of the funnel rather than embedded erupted material, with the caveat on floor-dominated regions noted above.

The eruptions are accompanied by an equatorial current sheet, across which the radial magnetic field reverses \citep{Ripperda2022}. To follow it we measure, at each radius and time, the fraction of the azimuth $f_{\rm cs}(r,t)$ in which such a reversal occurs inside the magnetically dominated region. At each $(r,\varphi)$ we take the Kerr--Schild radial field $B^r_\pm$ on the grid rows nearest $\theta = 90^\circ \mp 5^\circ$ and compute
\begin{equation}
c = -\frac{B^r_+ B^r_-}{\tfrac{1}{2}\left[(B^r_+)^2 + (B^r_-)^2\right]},
\label{eq:cs_reversal}
\end{equation}
which is $+1$ for equal and opposite fields and $-1$ for equal and parallel fields. A current sheet is counted where $c > 0.5$ and $\sigma > 1$. The magnetization cut excludes the turbulent disk field, which reverses across the midplane over about a third of the azimuth even between eruptions. Figure~\ref{fig:flux_eruption_overview}c shows $f_{\rm cs}$ over $t = 20{,}000$--$25{,}000~r_g/c$ at full resolution.
 
Between eruptions, the current sheet is confined to the inner magnetosphere: it covers most of the azimuth only within $r \lesssim 2\,r_g$, i.e.\ inside the equatorial ergosphere, and is almost absent further out. During each eruption the sheet grows outward and reaches its largest extent while the horizon flux is still falling, before the flux minimum. The strength of the eruption depends on how far it extends: to $\sim 2$--$3\,r_g$ in the three weaker eruptions and to $\approx 6\,r_g$ in the main one. These extents are lower limits, because the sheet can be displaced from the midplane where our probe does not see it. Over the full interval, the azimuthal coverage of the sheet tracks the rate at which the horizon loses flux and coincides with a reduced accretion rate (Section~\ref{sec:lagged}).

The main eruption shows the order of events. The sheet extends beyond the ergosphere shortly after the onset, grows to its largest extent, and then collapses back inside the ergosphere. Immediately afterwards the erupted bubbles reach their largest share of the jet, and the jet tilt roughly doubles (Figure~\ref{fig:flux_eruption_overview}c,e). The eruption thus proceeds from reconnection in an extended equatorial sheet, to the release of evacuated, magnetically dominated flux, to a strongly tilted jet. With only four eruptions we do not attempt a quantitative relation between sheet extent and eruption strength. This picture is consistent with plasmoid-mediated reconnection in an equatorial sheet extending $\sim 5$--$10\,r_g$ into the disk at $a = 0.9375$ \citep{Ripperda2022}, and with reconnection a few $r_g$ from the black hole at $a = 0.94$ \citep{Loules2026}.

For our case of $a = 0.998$ the horizon and the ISCO lie deep inside the equatorial ergosphere, where frame dragging forces the plasma and the field lines anchored in it to rotate with the black hole, and the toroidal field near the horizon becomes comparable to or stronger than the radial field, whereas it is negligible at $a = 0$ \citep{Zhang2024}. Between eruptions, the extent of the magnetically dominated sheet also marks where the magnetosphere meets the disk, which in our case lies inside the ergosphere. The sheet therefore forms in the region where frame dragging shapes the field, which governs the disk's angular momentum. Also, the jet region for the spinning case extends during the current sheet formation.

The radius-time diagrams of the non-cumulative angular momentum flux through the jet surface, $\dot{j}^\perp_{\rm jet\,surf}$ (Figure~\ref{fig:flux_eruption_overview}d), and of the mass flux, $\dot{M}(r,t)$ (Figure~\ref{fig:flux_eruption_overview}b), reveal the spatiotemporal structure of these events, complementing the three-dimensional view of the main eruption in Figure~\ref{fig:eruption_3d}. During flux eruptions, coherent structures propagate outward from the inner regions, visible as distinct features in the $r$--$t$ maps. The magnetic flux time series (Figure~\ref{fig:flux_eruption_overview}a) shows the episodic nature of the MAD state, with the flux oscillating around the saturation value $\phi_{\rm MAD} \approx 50$ but experiencing intermittent drops during eruption events, each preceded by the outward extension of the equatorial current sheet (Figure~\ref{fig:flux_eruption_overview}c). The jet morphology responds to these flux variations, with the tilt angle $\theta_{\rm tilt}$ and opening angle $\theta_{\rm open}$ showing correlated variability, both with and without the erupted bubbles (Figure~\ref{fig:flux_eruption_overview}e).

\subsection{Event Response Analysis}
\label{sec:event_response}

To quantify the impact of flux eruptions on the accretion-jet system 
(Figure~\ref{fig:eruption_3d}), we compare the state of key variables 
between the baseline and minimum windows defined in the previous 
subsection: the ``Event-start'' values are single-sample measurements at 
the onset time $t_{\rm on} = 23{,}430~r_g/c$, and the ``Event-end'' 
values are averages over the minimum window 
$t = 23{,}640$--$23{,}730~r_g/c$.
Table~\ref{tab:event_statistics} summarizes the changes in key quantities, quantifying the disruption visualized in Figure~\ref{fig:eruption_3d}.

\begin{deluxetable}{lrrr}
\tabletypesize{\footnotesize}
\tablecaption{Response of Key Quantities to the Main Flux Eruption Event
($t \approx 23{,}690~r_g/c$)\label{tab:event_statistics}}
\tablewidth{0pt}
\tablehead{
\colhead{Quantity} & \colhead{Start} & \colhead{End} & \colhead{Change (\%)}
}
\startdata
$\phi_{r_H}$ & 53.95 & 25.87 & $-52.0$ \\
$|\dot{M}|_{r_H}$ & 39.65 & 23.54 & $-40.6$ \\
$\dot{j}_{10}$ & 2.19 & $-0.90$ & reversal \\
$\theta_{\rm tilt}$ (deg) & 9.21 & 21.87 & $+137.5$ \\
\quad without bubble & 8.74 & 19.35 & $+121.3$ \\
$\theta_{\rm open}$ (deg) & 66.19 & 62.85 & $-5.0$ \\
\quad without bubble & 66.05 & 61.97 & $-6.2$ \\
\enddata
\tablecomments{``Start'' values are single-sample measurements at the
onset time $t_{\rm on} = 23{,}430~r_g/c$ (baseline window; see
Section~\ref{sec:eruption_id}); ``End'' values are averaged over the
minimum window $t = 23{,}640$--$23{,}730~r_g/c$. The $-52.0\%$ decrease
in $\phi_{r_H}$ is measured relative to this high-flux baseline and
therefore exceeds the 27.3\% detection depth, which is defined relative
to the local running mean. $|\dot{M}|_{r_H}$ is evaluated at the first
grid cell outside the horizon ($r = 1.07\,r_g$). $\dot{j}_{10}$ is the
non-cumulative angular-momentum flux through the jet surface averaged over
$9$--$11~r_g$, the outer edge of the radial range over which the jet angles
are averaged; it changes sign during the eruption, so no percentage is given.
The angles are computed for the instantaneous $\sigma > 1$ jet and, in the rows
marked ``without bubble'', after excluding the erupted bubble
($\delta_t < 0.01$; Section~\ref{sec:eruption_id}); they are averaged over
$r_H < r \le 10\,r_g$ (Appendix~\ref{app:jet_morphology_def}).}
\end{deluxetable}

The eruption is accompanied by coherent changes in all monitored quantities (Table~\ref{tab:event_statistics}). Shortly after the onset the equatorial current sheet extends beyond the ergosphere (Figure~\ref{fig:flux_eruption_overview}c), and both the horizon accretion rate and the horizon magnetic flux drop substantially. The angular-momentum flux through the jet surface at $10\,r_g$ does not merely decrease: it changes sign, so that around the flux minimum the net flux across the jet surface is briefly inward rather than outward (Figure~\ref{fig:flux_eruption_overview}d). The jet tilt more than doubles, and it does so whether or not the erupted bubble is counted as part of the jet, so the tilting is a property of the funnel itself. The opening angle responds in two stages (Figure~\ref{fig:flux_eruption_overview}e). It first widens, mostly because the bubble enters the magnetically dominated region; without the bubble the widening is small. It then contracts below its pre-eruption value. Because $\theta_{\rm open}$ measures the width of the $\sigma > 1$ region (Appendix~\ref{app:jet_morphology_def}), this contraction reflects the shrinking of the magnetically dominated funnel as horizon flux is expelled, rather than enhanced collimation of the outflow. This interpretation is supported by the positive correlation between $\phi_{r_H}$ and $\theta_{\rm open}$ (Table~\ref{tab:lagged_correlations}): the funnel width tracks the horizon flux that inflates it. These trends do not depend on the bubble threshold: for thresholds between $0.003$ and $0.1$ the tilt still more than doubles and the opening angle still ends below its pre-eruption value.

\begin{figure*}[ht!]
\centering
\includegraphics[width=\textwidth]{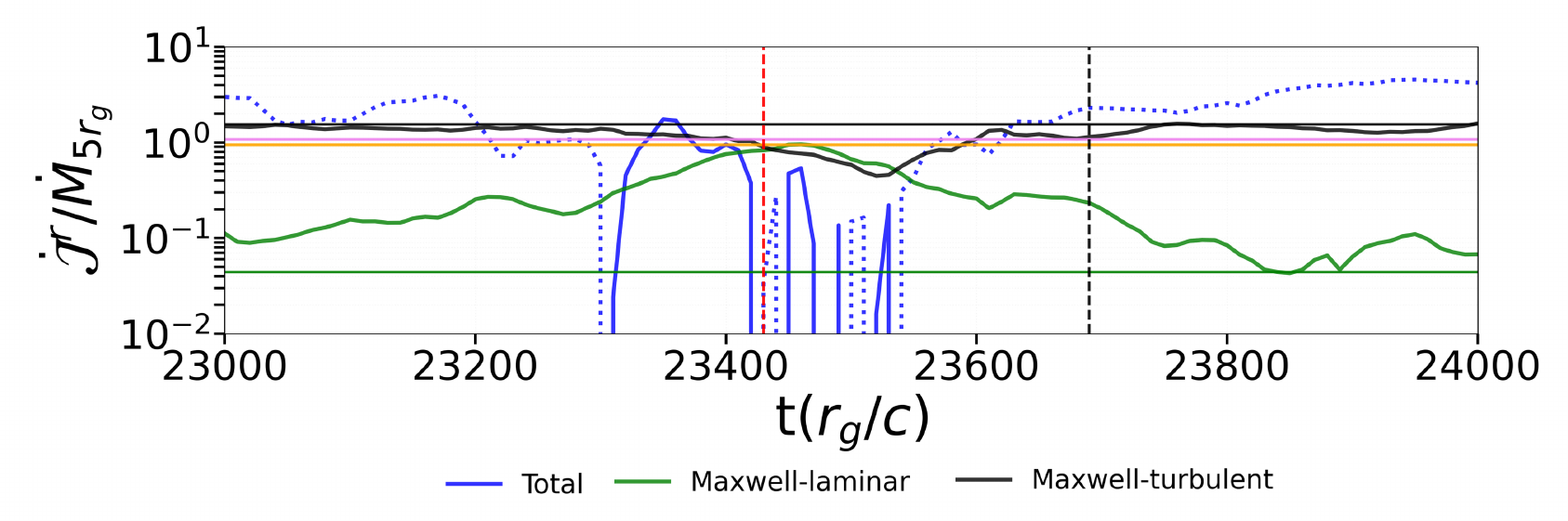}
\caption{Temporal evolution of radial angular momentum flux components in the disk region at $r = 20\,r_g$ during the main flux eruption event. The total flux (blue), laminar Maxwell stress (green), and turbulent Maxwell stress (black) are shown, with solid lines indicating positive (outward) flux and dotted lines indicating negative (inward) flux. Vertical dashed lines mark the eruption window ($t = 23{,}430$--$23{,}690~r_g/c$). Horizontal lines show time-averaged ($t = 20{,}000$--$25{,}000~r_g/c$) stress levels: green (MAD laminar Maxwell), black (MAD turbulent Maxwell), orange (INT turbulent Maxwell), and violet (SANE turbulent Maxwell).}
\label{fig:jdot_disk_temporal}
\end{figure*}

This response of the jet distinguishes eruptions around a rapidly spinning black hole from those in non-spinning MADs. At $a = 0$ there is no Blandford--Znajek jet, and the angular momentum carried by jets is nearly absent \citep{Zhang2024}, so eruptions act mainly on the disk, driving the winds and turbulent stresses that remove its angular momentum \citep{Chatterjee2022}. At $a = 0.998$ the same disk channel operates (Figure~\ref{fig:jdot_disk_temporal}), but the eruptions also interrupt a persistent, spin-powered jet: the angular-momentum flux through the jet surface at $10\,r_g$ reverses sign and the jet tilt more than doubles (Table~\ref{tab:event_statistics}). Because the jet-surface flux follows the horizon flux within $\sim 20\,r_g/c$ (Table~\ref{tab:lagged_correlations}), the larger fractional flux variability at high spin (Section~\ref{sec:eruption_id}) carries over into the jet.

To examine stress reorganization during the eruption, we analyze the radial angular-momentum-flux components in the disk at $r = 20\,r_g$ (Figure~\ref{fig:jdot_disk_temporal}). The laminar Maxwell component declines after the eruption onset, while the turbulent Maxwell component varies around values comparable to the time-averaged levels in the three runs before recovering. These local stress amplitudes demonstrate a temporary departure from the saturated MAD transport state, but do not establish a global magnetic-state transition. The expected phenomenological cycle is MAD $\rightarrow$ eruption $\rightarrow$ a lower-flux, potentially SANE-like transport phase $\rightarrow$ INT-like recovery $\rightarrow$ MAD; during the main event, however, the minimum horizon flux remains within the formal INT range.


\subsection{Lagged Correlation Analysis}
\label{sec:lagged}

To characterize the lead--lag relationships and propagation timescales in the system, 
we compute lagged Pearson correlation coefficients between pairs of time series. 
Table~\ref{tab:lagged_correlations} presents results for the full time series 
($t = 20{,}000$--$25{,}000~r_g/c$) which contains four identified flux-eruption events.

\begin{deluxetable}{lrrr}
\tabletypesize{\footnotesize}
\tablecaption{Lagged Correlation Analysis\label{tab:lagged_correlations}}
\tablewidth{0pt}
\tablehead{
\colhead{Variable pair} & \colhead{$r_0$} & \colhead{$r_{\max}$} &
\colhead{Lag}
}
\startdata
$-d\phi_{r_H}/dt$ vs $f_{\rm cs}$ & $+0.564$ & $+0.575$ & $-20$ \\
$\dot{M}_{r_H}$ vs $f_{\rm cs}$ & $-0.379$ & $-0.432$ & $-20$ \\
$\phi_{r_H}$ vs $\dot{M}_{r_H}$ & $+0.706$ & $+0.786$ & $-90$ \\
$\phi_{r_H}$ vs $\dot{j}_{10}$ & $+0.729$ & $+0.733$ & $+20$ \\
$\dot{M}_{r_H}$ vs $\dot{j}_{10}$ & $+0.464$ & $+0.594$ & $+60$ \\
$\phi_{r_H}$ vs $\theta_{\rm tilt}$ & $-0.580$ & $-0.627$ & $-60$ \\
\quad without bubble & $-0.550$ & $-0.573$ & $-50$ \\
$\phi_{r_H}$ vs $\theta_{\rm open}$ & $+0.354$ & $+0.728$ & $+170$ \\
\quad without bubble & $+0.512$ & $+0.759$ & $+140$ \\
\enddata
\tablecomments{$r_0$ is the Pearson correlation coefficient at zero lag;
$r_{\max}$ is the coefficient of largest magnitude for lags within
$\pm 200~r_g/c$, and Lag (in $r_g/c$) is the time offset at which it occurs.
A positive lag means that the second variable lags behind the first. The
lag window is comparable to the time from onset to flux minimum of the main
eruption and well below half the interval between eruptions, so the lags
describe the response within an eruption rather than the eruption cycle.
$f_{\rm cs}$ is the share of the azimuth with an equatorial current sheet
(Equation~\ref{eq:cs_reversal}) averaged over $2$--$5~r_g$, and
$-d\phi_{r_H}/dt$ is the rate of horizon-flux loss smoothed with a
$50~r_g/c$ running mean. $\dot{M}_{r_H}$ is evaluated at the first cell
outside the horizon, and $\dot{j}_{10}$ is the jet-surface flux averaged
over $9$--$11~r_g$. The angles refer to the instantaneous $\sigma > 1$ jet;
the rows marked ``without bubble'' exclude the erupted bubble
(Section~\ref{sec:eruption_id}).}
\end{deluxetable}

The lagged correlations (Table~\ref{tab:lagged_correlations}) follow the sequence of an eruption. The equatorial current sheet is extended while the horizon loses flux: its azimuthal coverage correlates with the rate of flux loss and is anticorrelated with the accretion rate, in both cases without measurable delay. The horizon flux and the accretion rate are tightly coupled, with the accretion rate changing slightly earlier. The angular-momentum flux through the jet surface at $10\,r_g$ follows the horizon flux within $\sim 20\,r_g/c$, comparable to the light-crossing time from the horizon, as expected for a flux carried by field lines anchored on the horizon. The jet tilt is anticorrelated with the horizon flux and is largest while the flux is still falling, whereas the opening angle lags the horizon flux by $\sim 150\,r_g/c$, consistent with the magnetically dominated funnel contracting only after the flux that supports it has been expelled. Excluding the erupted bubble leaves these relations essentially unchanged. The lags correspond to the maxima of broad correlation peaks and order the events only to within $\sim 100\,r_g/c$. Overall, the analysis supports coherent magnetospheric reconfiguration during eruptions.

\subsection{Physical Interpretation}
\label{sec:interpretation}

The results presented above paint a coherent picture of flux eruption dynamics 
in MADs around spinning black holes. The accumulated magnetic flux at the horizon ($\phi_{r_H} \sim 50$--$55$) 
maintains an organized, large-scale poloidal field structure that efficiently 
supports a strong outward electromagnetic angular-momentum flux through laminar Maxwell stresses
(Figure~\ref{fig:jdot_jetsurf_profiles}). 

When magnetic interchange instabilities trigger a flux eruption, the organized 
field structure is disrupted. The angular momentum affected by the eruptions has two sources (Section~\ref{sec:flux_transport}): the jet flux is drawn from the black-hole spin, whereas the eruption-driven winds and stresses redistribute the disk's own angular momentum. Compared with non-spinning MADs \citep{Chatterjee2022, Narayan2022, Zhang2024}, eruptions around the rapidly spinning black hole occur at a similar saturated flux but with larger fractional flux variability; they interrupt a Blandford-Znajek jet in addition to driving the disk winds and stresses, so they modulate the outflow of the black hole's own angular momentum; and they are associated with a current sheet that is confined to the ergosphere between eruptions and extends beyond it during them.

The characteristic timescales of the eruption cycle---the eruption durations ($\sim 40$--$90\,r_g/c$) and the lags between the horizon flux, accretion rate and jet tilt ($\sim 50$--$90\,r_g/c$; Table~\ref{tab:lagged_correlations})---correspond to the
reconfiguration of the inner accretion flow and jet base. This rapid, sub-viscous recovery is regulated by the erupted
flux tube itself, which is sheared apart before reaching the disk's
pressure maximum; beyond the pressure maximum, flux reaccumulation instead proceeds on the viscous timescale via the advection--diffusion balance of the large-scale field
\citep{JacqueminIde2026}. For a stellar-mass black hole
of $10~M_\odot$, these timescales translate to $\sim 2$--$4$~ms. For Sgr~A$^*$, whose black-hole mass is $M \sim 4 \times 10^6~M_\odot$, the corresponding
timescales are $\sim 15$--$30$~minutes, comparable to the quasi-periodic 
variability observed in near-infrared and millimeter wavelengths 
\citep{GRAVITY2018, Wielgus2022}.

The jet wobbling during flux eruptions, with tilt variations of 
$\sim 10^\circ$--$20^\circ$ on timescales of hundreds of $r_g/c$ actively distorts the jet bulk \citep{Nalewajko:2026, Davelaar2023} and could be relevant to the jet position-angle variability observed in VLBI monitoring of AGN jets 
\citep{Lister2021}. The association between flux state and jet geometry, with the opening angle lagging the horizon flux by $\sim 140$--$170\,r_g/c$
(Table~\ref{tab:lagged_correlations}), suggests that changes in the 
magnetic-flux state could produce observable changes in jet morphology 
and orientation stability \citep{Chatterjee:2025}.

\section{\label{sec:conclusion}Summary}

Our analysis reveals a progression in the net angular-momentum flux from SANE through INT to MAD. SANE is advection dominated and retains a net inward flux, whereas the MAD run is dominated by a coherent outward electromagnetic contribution with an approximately constant radial profile; INT remains transitional. Comparison with the non-spinning systems of \citet{Chatterjee2022} 
suggests that the Blandford--Znajek torque introduces an additional 
angular-momentum source in the spinning MAD, shifting the global budget 
from net inward to net outward flux. This net outward flux combines two 
channels with distinct sources: the jet drains the black hole's spin 
angular momentum via the Blandford--Znajek torque, whereas the disk's 
angular momentum is removed by the internal disk stresses and by the 
wind and disk-surface fluxes. The jet-dominated positive total flux is 
therefore a signature of spin-energy extraction, not by itself a measure 
of disk transport efficiency.

Analysis of angular momentum flux through jet--wind--disk boundaries reveals distinct coupling mechanisms. MAD exhibits efficient magnetic braking at the wind--jet interface through organized large-scale fields and positive flux through the disk surface, indicating net vertical extraction. In contrast, SANE shows negligible jet--wind coupling and anomalous behavior where angular momentum flows \textit{into} rather than out of the disk through vertical boundaries. The MAD state exhibits episodic flux eruption events that produce coordinated suppression of mass accretion and angular-momentum flux and modify the jet morphology, with characteristic timescales of $\sim 40-90\,r_g/c$. Lagged correlations show close coupling between magnetic flux, accretion 
rate, and jet morphology, with the jet opening angle lagging the horizon flux by $\sim 140$--$170\,r_g/c$, consistent with coherent magnetospheric reconfiguration during eruptions. Each eruption is preceded by the extension of the equatorial current sheet from inside the ergosphere to $\sim 2$--$3\,r_g$ in weaker and $\approx 6\,r_g$ in the strongest eruptions, and is followed by low-density magnetized bubbles that carry up to $31\%$ of the magnetization-weighted jet volume within $10\,r_g$. Excluding these bubbles reduces the jet tilt at the flux minimum from $21.9^\circ$ to $19.4^\circ$ but leaves the more-than-doubling of the tilt and the late contraction of the opening angle unchanged. Compared with non-spinning MADs \citep{Chatterjee2022, Narayan2022, Zhang2024}, eruptions around the rapidly spinning black hole occur at a similar saturated flux but with larger fractional flux variability; they interrupt a spin-powered jet in addition to driving the disk winds and stresses, so they modulate the outflow of the black hole's own angular momentum; and they are associated with a current sheet that is confined to the ergosphere between eruptions and extends beyond it during them.

The stress-to-pressure ratio, increases with magnetization at all radii. Within $\sim2\,r_g$ it is dominated by Maxwell stress in all three runs, with a laminar share of $50$--$90$ per cent at the horizon that increases with magnetic flux; near the horizon it should therefore be read as a measure of magnetic torque acting on a plunging flow rather than as a turbulent viscosity. In the bulk disk the SANE and INT stresses are turbulent, with Reynolds stress contributing $30$--$100$ per cent of the Maxwell stress, whereas MAD retains a $\sim25$--$30$ per cent laminar Maxwell component and a weak inward Reynolds stress. The corresponding viscous timescales over $r=10$--$40\,r_g$ are $\sim0.15$--$5$~s for stellar-mass black holes and hours to days for supermassive black holes, extrapolating to months--years at $r\sim10^2$--$10^3\,r_g$, consistent with the transition timescales of changing-look AGN. MAD evolves $\sim3.5$--$4.5$ times faster than SANE at fixed radius.

The measured eruption timescales scale to milliseconds for stellar-mass black holes and minutes for Sgr~A$^*$. The simulated jet wobble ($\Delta\theta_{\rm tilt} \sim 10^\circ$--$20^\circ$) also motivates comparison with position-angle variability in VLBI monitoring, although radiative transfer and propagation to observable scales are required for a direct prediction.

Magnetic-flux saturation emerges as a key parameter controlling the angular-momentum budget in these rapidly rotating black-hole systems. In the MAD run, organized fields connect the horizon, jet, wind, and disk and support a jet-dominated outward flux; in SANE, inward advection and turbulent disk stresses remain more important. This framework links the magnetic-flux state to accretion dynamics and jet variability.

The interpretation is limited by the use of a single black-hole spin, one realization of each magnetic configuration, ideal-GRMHD dissipation, a finite averaging interval, and only four identified eruption events, which limit the robustness of the lagged-correlation analysis and of the relation between current-sheet extent and eruption strength. Our density-based identification of erupted bubbles cannot detect erupted flux inside the floor-dominated polar funnel, so the bubble contribution to the jet is a lower limit. The observational implications are dimensional scalings because the simulations do not include radiation or synthetic observables.

\begin{acknowledgments}
R.R. acknowledges the Prime Minister’s Research Fellowship (PMRF) scheme for providing fellowship. 
\end{acknowledgments}

\appendix

\section{Component Analysis of Surface Angular Momentum Fluxes}
\label{app:surface_flux}

\begin{figure*}[ht!]
\centering
\includegraphics[width=\textwidth]{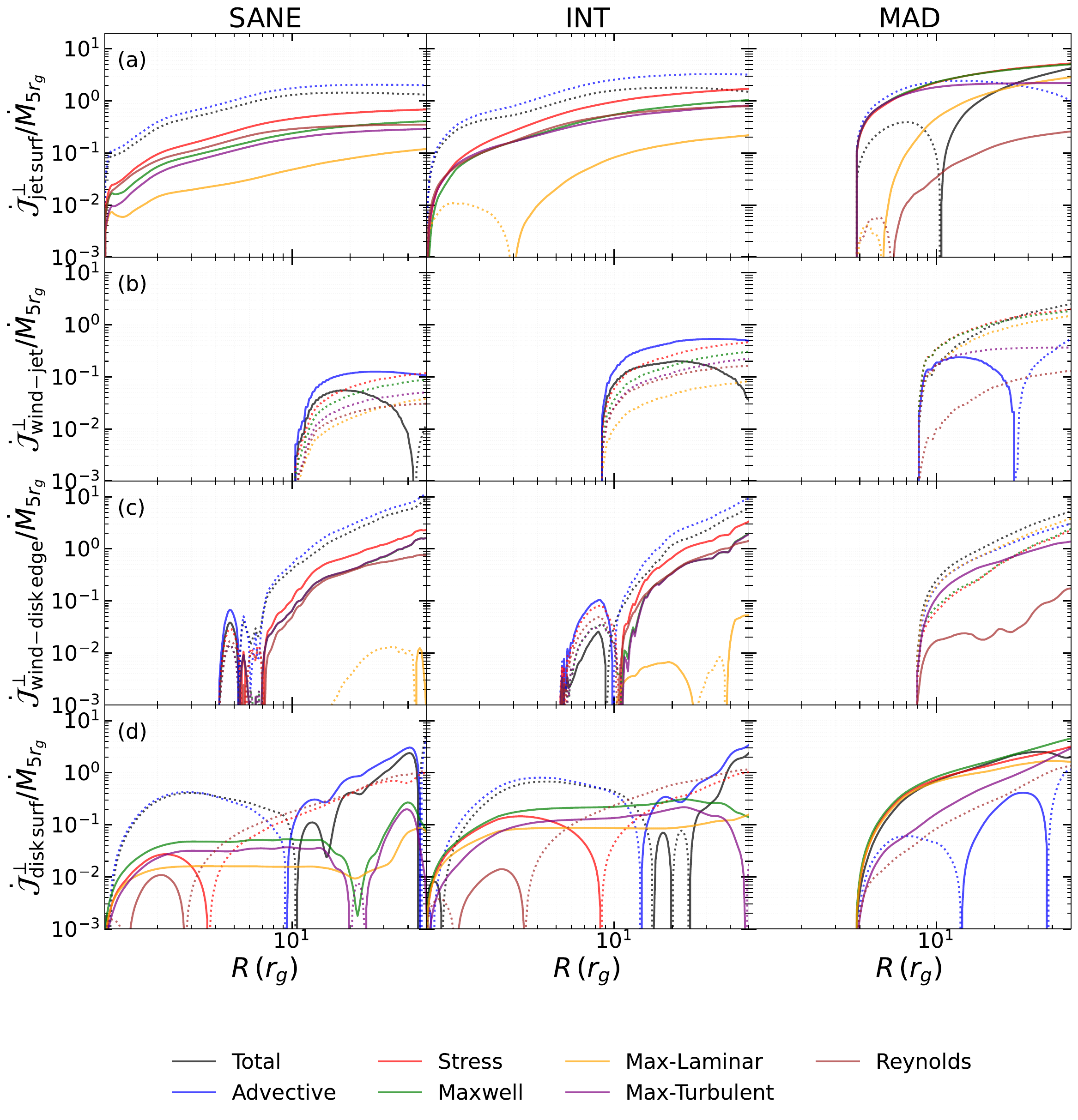}

\caption{Cumulative angular momentum flux through key surfaces for SANE, 
INT, and MAD configurations. Solid lines indicate positive flux 
(outward transport: for jet surface, from jet to wind/disk; for wind--jet surface, 
from wind to jet; for wind--disk edge surface, from wind to disk edge; for disk 
surface, from disk to jet/wind/disk-edge); dotted lines indicate negative flux 
(inward transport). Fluxes are normalized by $\dot{M}_{5r_g}$. 
(a)~\textit{Jet surface}: MAD transitions from negative at $r \lesssim 10\,r_g$ 
to positive at large radii, reflecting competition between advective transport 
(consistently negative, carrying angular momentum into the jet) and stress-driven 
extraction (consistently positive); INT and SANE show persistently negative flux 
throughout, with advection dominating and positive stress insufficient to reverse 
net inward transport. 
(b)~\textit{Wind--jet interface}: MAD shows strongly negative flux indicating 
substantial angular momentum transfer from jet to wind, increasing with radius; 
INT and SANE exhibit near-zero net flux with weak positive advection balanced by 
weak negative stress. 
(c)~\textit{Wind--disk edge interface}: all three configurations show negative 
flux indicating angular momentum transfer from disk edge to wind. 
(d)~\textit{Disk surface}: MAD extraction driven by Maxwell stress (laminar and 
turbulent, both positive); INT shows inward transport at inner radii from advective 
torque with oscillatory behavior, transitioning at outer radii to competition 
between outward advective and inward Reynolds stress; SANE exhibits anomalous 
behavior with negative flux dominated by advective torque at intermediate radii, 
strongly positive at large radii, and strongly negative again around 
$\sim 50\,r_g$; laminar and turbulent Maxwell stress are always positive but 
subdominant in INT and SANE.}
\label{fig:jdot_surfaces}
\end{figure*}

Here we provide detailed component decompositions of the cumulative perpendicular 
angular momentum flux through the four key surfaces shown in 
Figure~\ref{fig:jdot_surfaces}. We note that the wind--jet flux is not simply the 
complement of the jet surface flux: at small radii ($r \lesssim 10\,r_g$), the 
jet boundary may directly contact the disk region before a well-defined wind zone 
develops, so the two profiles have different radial onsets.

\paragraph{Jet Surface.}
In MAD, both the laminar and turbulent Maxwell components contribute comparably to 
the net angular momentum extraction through the jet surface, with the turbulent 
Maxwell stress playing a particularly prominent role at the base of the jet. The 
Reynolds stress provides a minor but consistently positive contribution throughout. 
In INT, the Reynolds stress and turbulent Maxwell stress each contribute a major 
and roughly equal share of the total torque transporting angular momentum out of 
the jet, while the laminar Maxwell contribution is substantially weaker than in 
MAD. In SANE, the Reynolds stress becomes the dominant component of the total 
stress, with turbulent Maxwell playing a secondary role and laminar Maxwell stress 
negligible, consistent with the absence of organized large-scale field structure.

\paragraph{Wind--Jet Interface.}
In MAD, the magnetic braking is dominated by strongly negative laminar Maxwell 
stress, reflecting the organized large-scale magnetic field that threads both the 
jet and wind regions. The turbulent Maxwell and Reynolds stresses provide smaller 
but consistently negative contributions. The advective torque transitions from 
positive at small radii (opposing the magnetic braking) to negative at large radii 
(assisting it), indicating that bulk flow eventually reinforces the magnetic 
angular momentum transfer from jet to wind. In INT, the advective torque initially 
transports angular momentum from wind to jet, but it weakens with radius and is 
eventually dominated by the stress component, which drives the flux from jet to 
wind. A qualitatively similar progression occurs in SANE, but the transition from 
advection-dominated to stress-dominated transport occurs at smaller radii. In both 
INT and SANE, the net flux is near-zero throughout, with weak positive advection 
approximately balanced by weak negative stress, resulting in negligible net 
wind--jet coupling.

\paragraph{Wind--Disk Edge Interface.}
In MAD, both the advective and total stress torques act to transport angular 
momentum from the disk edge into the wind, resulting in persistently negative 
cumulative flux. Notably, the laminar Maxwell stress is strongly negative while 
the turbulent Maxwell stress is positive, indicating an internal competition where 
organized large-scale fields extract angular momentum from the disk edge but 
turbulent field fluctuations partially counteract this extraction. There is, 
however, a minor positive (wind-to-disk-edge) flux contribution at inner radii 
driven by the advective torque. In INT and SANE, the magnitude of the 
disk-edge-to-wind angular momentum transfer is even larger than in MAD, but it 
is driven almost entirely by advection, with the stress component acting in the 
opposite (positive) direction. The laminar Maxwell stress is essentially zero in 
SANE, confirming the complete absence of coherent large-scale field contributions, 
while the turbulent Maxwell stress---though present---acts to oppose rather than 
reinforce the advective extraction.

\paragraph{Disk Surface.}
In MAD, the disk surface flux is positive throughout, indicating net angular 
momentum extraction from the disk through its upper and lower boundaries. This 
extraction is driven by Maxwell stress, with both the laminar and turbulent 
components positive and comparable in magnitude. In INT, the flux is negative 
(inward transport) at inner radii, dominated by the advective torque, with 
oscillatory behavior reflecting the variable magnetization state. At outer radii, 
the flux is governed by competition between outward advective torque and inward 
Reynolds stress. SANE exhibits more complex behavior: the flux is negative and 
advection-dominated at intermediate radii (similar to INT), becomes strongly 
positive at large radii, and then turns strongly negative again around 
$r \sim 50\,r_g$. In both INT and SANE, the laminar and turbulent Maxwell stresses 
remain positive throughout but are subdominant, unable to overcome the advective 
and Reynolds stress contributions. The Reynolds stress is negative across all 
three configurations at the disk surface, opposing magnetic extraction in a manner 
consistent with turbulent mixing.

\section{Jet Morphology: Tilt and Opening Angles}
\label{app:jet_morphology_def}

To characterize the jet geometry and its response to flux eruptions, we compute 
two morphological parameters: the tilt angle $\theta_{\rm tilt}$ and the opening 
angle $\theta_{\rm open}$.

The tilt angle measures the misalignment between the jet axis and the black hole spin axis. In each snapshot we take the jet to be the instantaneous region with $\sigma = b^2/\rho > 1$ and assign each jet cell to the upper or lower jet by the sign of $b^r$ for $r \le 25\,r_g$ and by the sign of $z$ at larger radii. For each jet and each radial shell we compute the magnetization-weighted centroid
\begin{equation}
\mathbf{D}_{\rm jet}(r) = \frac{\int_{\rm jet} \sigma \, \mathbf{x} \, \sqrt{-g}\, d\theta\, d\varphi}{\int_{\rm jet} \sigma \, \sqrt{-g}\, d\theta\, d\varphi},
\label{eq:jet_direction}
\end{equation}
where $\mathbf{x}$ is the Cartesian position vector and the integrals run over the jet cells in that shell. The tilt angle is
\begin{equation}
\theta_{\rm tilt}(r) = \arccos\left(\frac{|\mathbf{D}_{\rm jet} \cdot \hat{\mathbf{z}}|}{|\mathbf{D}_{\rm jet}|}\right),
\label{eq:tilt_angle}
\end{equation}
where $\hat{\mathbf{z}}$ is the unit vector along the black hole spin axis. A tilt angle of zero indicates perfect alignment, while nonzero values indicate jet misalignment or wobbling.
 
The opening angle quantifies the angular width of the jet. It is computed from the volume-weighted mean distance of the jet cells from the centroid of the same shell,
\begin{equation}
\langle \Delta \rangle(r) = \frac{\int_{\rm jet} |\mathbf{x} - \mathbf{D}_{\rm jet}| \, \sqrt{-g}\, d\theta\, d\varphi}{\int_{\rm jet} \sqrt{-g}\, d\theta\, d\varphi},
\end{equation}
as
\begin{equation}
\theta_{\rm open}(r) = \frac{3}{2} \frac{\langle \Delta \rangle}{r}.
\label{eq:opening_angle}
\end{equation}
The factor $3/2$ converts the mean distance into the radius of a uniformly filled circular cross-section, for which the mean distance from the centre is two-thirds of the radius, so $\theta_{\rm open}$ is an effective half-opening angle. During flux eruptions the $\sigma = 1$ boundary itself evolves, so $\theta_{\rm open}$ conflates changes in the outflow geometry with changes in funnel magnetization; it should therefore be interpreted as the angular width of the magnetically dominated region rather than of the mass-loaded jet.
 
The values quoted in the text, Tables~\ref{tab:event_statistics} and \ref{tab:lagged_correlations}, and Figure~\ref{fig:flux_eruption_overview} are averages over $r_H < r \le 10\,r_g$ and over the two jets, computed every $10\,r_g/c$ with the data coarsened by a factor of 4 in $\theta$ and $\varphi$. For the ``jet without bubble'' values, cells of the erupted bubble ($\delta_t < 0.01$; Section~\ref{sec:eruption_id}) are removed from the jet before the same calculation.
 
In the MAD state, the time-averaged values over $t = 20{,}000$--$25{,}000~r_g/c$ are $\langle \theta_{\rm tilt} \rangle = 6.9^\circ \pm 3.8^\circ$ and $\langle \theta_{\rm open} \rangle = 63.6^\circ \pm 1.7^\circ$ ($6.6^\circ \pm 3.3^\circ$ and $63.5^\circ \pm 1.5^\circ$ without the bubble). During flux eruption events, these angles vary as the magnetic field reconfigures during eruptions.

\bibliography{ref}
\bibliographystyle{aasjournalv7}

\end{document}